\documentclass[aps,prb,floatfix,onecolumn,showpacs,superscriptaddress,longbibliography]{revtex4-2}
\usepackage{mathtools,amssymb,graphicx}
\usepackage{soul} 
\usepackage[dvipsnames,usenames]{color}
\usepackage[normalem]{ulem}
\usepackage{tabularx}
\usepackage{graphicx}
\usepackage{xcolor}
\usepackage{bbold, verbatim} 
\usepackage{float}
\usepackage{tabularx}
\usepackage{rotating}
\usepackage{braket}
\usepackage{multirow}
\usepackage{ulem}
\usepackage{rotating}

\usepackage{inconsolata} 
\usepackage{listings}
\lstdefinestyle{pseudocode}{
  basicstyle=\ttfamily\small,
  numbers=left,
  numbersep=8pt,
  frame=lines,
  columns=fullflexible,
  showstringspaces=false,
  keepspaces=true,
  upquote=true,
  tabsize=2
}
\begin{document} 
\title{Fast Evaluation of Unbiased Atomic Forces in ab initio Variational Monte Carlo via the Lagrangian Technique}
\author{Kousuke Nakano}
\email{kousuke\_1123@icloud.com}
\affiliation{Center for Basic Research on Materials (CBRM), National Institute for Materials Science (NIMS), 1-2-1 Sengen, Tsukuba, Ibaraki 305-0047, Japan}
\author{Stefano Battaglia}
\email{stefano.battaglia@chem.uzh.ch}
\affiliation{Department of Chemistry, University of Zurich (UZH), Winterthurerstrasse 190, 8057, Zurich, Switzerland}
\author{J\"{u}rg Hutter}
\affiliation{Department of Chemistry, University of Zurich (UZH), Winterthurerstrasse 190, 8057, Zurich, Switzerland}
\date{\today}

\begin{abstract}
Ab initio quantum Monte Carlo (QMC) methods are state-of-the-art electronic structure calculations based on highly parallelizable stochastic frameworks for accurate solutions of the many-body Schr{\"o}dinger equation, suitable for modern many-core supercomputer architectures. Despite its potential, one of the major drawbacks that still hinders QMC applications, especially when targeting dynamical properties of large systems or extensive datasets, is the lack of an affordable method to compute atomic forces that are consistent with the corresponding potential energy surfaces (PESs), also known as \emph{unbiased} atomic forces. Recently, one of the authors in the present paper proposed a way to obtain unbiased forces with the Jastrow-correlated Slater determinant ansatz, where the determinant part is frozen to the values obtained by a mean-field method, such as Density Functional Theory [{\it Phys. Rev. B} \underline{109}, 205151 (2024)]. However, the proposed method has a significant drawback for its applications: for a system with $N$ nuclei, one requires 6$N$ additional DFT calculations to get unbiased forces, which is not negligible as the system size increases. This paper presents a way to replace the 6$N$ DFT calculations with a single coupled-perturbed Kohn-Sham calculation, following the so-called {\emph{Lagrangian technique}} established in quantum chemistry. This improves the computational cost and scalability of the method. We also demonstrate that the developed unbiased VMC force calculation improves not only the consistency with PESs, but also its accuracy, by investigating three molecules from the rMD17 benchmark set, and comparing the unbiased VMC forces with those obtained by the Coupled-Cluster Singles and Doubles with perturbative Triples [CCSD(T)] calculations. We found that the bare VMC forces are biased from the CCSD(T) ones, while the unbiased ones give values closer to those of the CCSD(T) ones. Our benchmark test also reveals that the unbiased VMC forces yield very consistent values with hybrid and meta GGAs (e.g., $\omega$B97X-D3BJ and $\omega$B97M-D3BJ), but do not necessarily yield values that are very close to those of CCSD(T). Our finding paves the way to generate machine learning interatomic potentials based on VMC forces more efficiently and accurately.
\end{abstract}
\maketitle

\makeatletter
\def\Hline{
\noalign{\ifnum0=`}\fi\hrule \@height 1pt \futurelet
\reserved@a\@xhline}
\makeatother

\section{Introduction}
\label{sec:intro}
{\vspace{2mm}}
Ab initio quantum Monte Carlo (QMC) methods have been studied as accurate and highly parallelizable electronic structure methods that can leverage the capabilities of current and near-future exascale supercomputers~{\cite{shinde2025shifting}}. While density functional theory (DFT) is the most widely used ab initio method, it sometimes faces significant challenges in accurately describing peculiar systems, such as strongly-correlated and dispersion-dominated materials. QMC methods are attracting attention as promising alternatives to address these limitations.
Despite the maturity of energy calculations by QMC, the atomic force calculation (i.e., first derivatives of the total energy with respect to atomic positions) has remained an open challenge, limiting the broader applicability of QMC. There are two main real-space QMC frameworks: variational Monte Carlo (VMC) and fixed-node diffusion Monte Carlo (FN-DMC)~{\cite{2001FOU}}. In this study, we focus on VMC because the calculation of forces within the FN-DMC framework~{\cite{Flaviano2025_DMC_comparison}} is much more complicated. The FN-DMC force calculation is challenging because the FN wavefunction is not an eigenstate for the force estimator unlike for the Hamiltonian (c.f., therefore, forces are computed using the so-called mixed-estimator) and the derivative of the FN wavefunction is hard to compute (c.f., it is typically replaced with the trial WF)~{\cite{1986REY, 2005CHI_Chiesa_scheme, 2008BAD2, 2011ASS, 2014MOR, 2021VAN}}. Notice that other QMC frameworks, such as auxiliary-field quantum Monte Carlo (AFQMC) approach, can also compute atomic forces and pressures~{\cite{2023AFQMC-Force}}.
The methodological development for force evaluation in VMC has a long history, dating back to the 1980s. An \emph{intrinsic} difficulty in computing atomic forces by VMC is the so-called {\it infinite-variance problem}: when the force is evaluated by Monte Carlo sampling, the variance of the force estimator diverges~\cite{2000ASS_VMC_infinite_variance}. Over the past decades, this difficulty has been addressed by several complementary strategies. One of the earliest attempts was the space-warp coordinate transformation (SWCT) originally introduced by Umrigar~{\cite{Umrigar1989_SWCT}}. This was generalized by Filippi and Umrigar within a correlated sampling framework~\cite{Filippi2000CorrelatedSamplingForces}, and then, its differential formulations were developed by Assaraf and Caffarel~{\cite{Assaraf2003_ZVZB_force}} and Sorella and Capriotti{~\cite{Sorella2010_AAD}}. Although these advances cure the divergence problem associated with the electron–ion Coulomb singularity, the other divergence arising from the nodal surface also had to be addressed~\cite{Attaccalite2008_AS}. Successful remedies are reweighting of the sampling distribution~\cite{Attaccalite2008_AS,Pathak2008_PW_reweight} and the tail regression estimator~\cite{Trail2008HeavyTails,Trail2008AltSampling,LopezRios2019TailRegression}.  
Related to the divergence problem, it was later revealed that one must also consider linear dependence of localized Gaussian basis sets (i.e., ill-conditioned overlap matrices) in periodic boundary condition calculations ~\cite{Nakano2021PRB_finite_variance_of_forces_for_PBC}.
From an implementation point of view, the application of the adjoint algorithmic differentiation to atomic force evaluation~{\cite{Sorella2010_AAD}} was a major milestone. It allows one to compute atomic forces without hand-coded analytic derivatives, dramatically reducing human effort in the derivative implementation~{\cite {Sorella2010_AAD}}. Subsequently, a fast scheme for derivatives of multi-determinant wave functions was also formulated~\cite{Filippi2016_fastderivatives_multideterminant}, which has recently facilitated the generation of high-quality training data for machine learning force fields~\cite{slootman2024accurate}.
Very recently, neural network ansatze combined with the VMC approach have been proposed to compute accurate atomic forces and pressures~{\cite{qian2022interatomic_force_nnvmc, qian2024nn-force-solids, lai2024qian-kunnet-force}}.
These developments have brought VMC force evaluations closer to practical utility.

{\vspace{2mm}}
Nevertheless, force calculations in VMC have been limited to small molecules~{\cite{2005CHI_Chiesa_scheme,2013ZEN_water,2014MOR,2014LUO_QMC_MD,2019LIU_qmc_force_onstant,2021VAN, Tiihonen2021JCP, Nakano2022JCP,slootman2024accurate}}, clusters~{\cite{2017MOU, mouhat2023thermal}}, and crystals with light elements such as solid/liquid hydrogen~{\cite{2018MAZ_Hydrogen, 2022TIR_MLP_hydrogen, 2022LY_qmc_hydrogen_phonon, 2023NIU_qmc_hydrogen, 2024CEP_qmc_hydrogen, 2025Giacomo, Kevin2025_hydrogen_RQMC, Goswami2025_hydrogen_MLP}} and diamond~{\cite{Nakano2021PRB_finite_variance_of_forces_for_PBC}}. This limitation likely comes from a {\it practical} issue underlying VMC-based force evaluations, which we will describe hereafter:
Let ${\bf R}_{\alpha}$ be the atomic position of the nucleus $\alpha$. 
The atomic force acting on $\alpha$ is defined as the negative gradient of the energy with respect to ${\bf R}_{\alpha}$:
\begin{subequations}
\begin{align}
{\bf F}_{\alpha}^\text{VMC} = - \frac{{\rm d} E}{{\rm d}{\bf R}_{\alpha}} = &- \Braket{\frac{\partial}{\partial {\bf R}_{\alpha}} E_{\rm L}} \label{eq:hf} \\
&-2\Braket{(E_{\rm L} - E) \frac{\partial \log \Psi}{\partial {\bf R}_{\alpha}}}\label{eq:pulay} \\
&-\sum_{i=1}^{N_p}{\frac{\partial E}{\partial p_{i}}} {\frac{d p_{i}}{dm {\bf R}_{\alpha}}}, \label{eq:add}
\end{align}
\end{subequations}
where $\Psi$ is an optimized wavefunction, $\braket{A}$ indicates the quantum average of the local operator $A$ over the sampling of $|\Psi|^2$, $E_{\rm L}$ is the so-called local energy ($E_{\rm L} \equiv \hat{H}\Psi/\Psi$), with $E \equiv \braket{E_L}$, and $\{ p_1, \cdots, p_{N_{p}}\}$ is the set of $N_{p}$ variational parameters included in the $\Psi$ ansatz. Eqs.~(\ref{eq:hf}), (\ref{eq:pulay}), and (\ref{eq:add}) are called the Hellmann--Feynman (HF), Pulay, and non-variational (NV) terms, respectively. When the wavefunction includes all relevant variational parameters and is fully optimized, the NV term vanishes~{\cite{Tiihonen2021JCP, Nakano2024}}. However, especially for large systems or extensive datasets, optimization of all wavefunction variational parameters is often infeasible in practice. Therefore, the so-called Jastrow-Slater determinant approach is typically employed, where the Slater determinant {\emph{obtained from mean-field methods}} such as DFT is kept fixed, and only the Jastrow factor is variationally optimized. Historically, even in this case, the NV term was often neglected on the assumption that its contribution was small, and forces were computed using only the HF and Pulay terms.
However, recent studies have demonstrated that ignoring the NV term can lead to significant biases~{\cite{Tiihonen2021JCP, Nakano2022JCP, Nakano2024}}. The bias arising from the NV term is referred to as the self-consistency error (SCE)~{\cite{Tiihonen2021JCP}}. In 2021, Tiihonen et al.~{\cite{Tiihonen2021JCP}} reported that the SCE becomes more severe for heavier elements. Follow-up studies confirmed that the SCE for liquid hydrogen is on the order of ~1 GPa~{\cite{2024GIA_qmc_hugonioit}}, and it reaches up to 5 GPa for cubic boron nitride (c-BN)~{\cite{Nakano2024}}. This implies that applying VMC to realistic systems beyond model compounds should address the SCE issue; otherwise, it becomes a critical bottleneck in VMC applications.

{\vspace{2mm}}
Nakano et al. (one of the authors of this present paper) recently formulated a method to eliminate the SCE~{\cite{Nakano2024}}. The idea is conceptually straightforward: a direct evaluation of the NV term using a hybrid VMC-DFT approach. The partial derivatives $\partial E / \partial p_{i}$ are computed via standard VMC, while the wavefunction response term $dp_i/d{\bf R}_{\alpha}$ is evaluated using DFT by the finite difference method (FDM). Using this scheme, unbiased forces have been successfully obtained for various systems such as H$_2$, Cl$_2$, c-BN~{\cite{Nakano2024}}, and liquid hydrogen~{\cite{2025Giacomo}}.
Although this methodology successfully demonstrated a proof-of-concept for SCE elimination~{\cite{Nakano2024, 2025Giacomo}}, a significant practical limitation still remains. The problem is that \emph{6$N$ additional DFT calculations are required} to compute unbiased forces on all $N$ atoms~{\cite{Nakano2024}}. While this is acceptable for small systems, it becomes computationally prohibitive when targeting large systems or preparing extensive datasets for machine-learning potential (MLP) training.

{\vspace{2mm}}
This paper presents a method to eliminate the 6$N$ additional DFT calculations when correcting the NV term to get unbiased VMC forces. The key idea is the following: The bias arising from the NV term is not a specific problem in VMC calculations, but also in wavefunction theory (WFT) calculations when non-variational wavefunctions, such as those used in perturbative methods, are used to compute forces. One well-established approach in WFT is the so-called {\emph{Lagrangian technique}}~{\cite{helgaker1989configuration, 2017Jensen_textbook}}. In this work, we extend the Lagrangian technique to the VMC framework, enabling the unbiased atomic force and pressure calculations for both molecular and periodic systems by a single-shot VMC and a single-shot coupled-perturbed Hartree-Fock (CPHF) or coupled-perturbed Kohn-Sham (CPKS) calculation. This eliminates the need for 6$N$ DFT evaluations and improves the feasibility of the method.
Furthermore, we demonstrate that the developed unbiased force evaluation improves not only the consistency, but also the accuracy of the VMC forces with the Jastrow Slater determinant ansatz. We benchmarked our method on three molecules, ethanol, malonaldehyde and benzene, taken from rMD17 benchmark set~{\cite{Christensen2020_MLST, Christensen2020_rMD17_Figshare}, and compare biased and unbiased VMC forces to those obtained with the Coupled-Cluster Singles and Doubles with perturbative Triples [CCSD(T)]~{\cite{Raghavachari1989_CCSDT}} method, known as the gold-standard in quantum chemistry calculations.

{\vspace{2mm}}
This paper is organized as follows. In Sec.~{\ref{sec:method}} we describe the details of the Lagrangian technique for the VMC framework. In Sec.~{\ref{sec:validation}}, we show the validation of our formalism and implementations in CP2K~{\cite{Kuehne2020_CP2K_JCP}} and TurboRVB~{\cite{2020NAK2}} software suites. In Sec.~{\ref{sec:application}}, we report benchmark results of the unbiased VMC force calculations for the three molecules taken from the rMD17 benchmark set. In Sec.~{\ref{sec:concluding-remarks}}, we conclude our work and discuss the current limitations and future works of the developed method.

\section{Methods: Lagrangian technique extended to VMC}
\label{sec:method}
The key idea of the Lagrangian technique~{\cite{helgaker1989configuration, 2017Jensen_textbook}} is to construct a function, \emph{the Lagrangian}, that has the same energy as the VMC one, but that incorporates all the constraints imposed during the optimization of the wavefunction. By making the Lagrangian stationary with respect to all its parameters, we then obtain a fully variational expression for which we can compute the nuclear gradient directly.
The VMC energy depends on a set of Jastrow variational parameters $\{p_i\}_{i=1}^{N_J}$, collected in the vector $\mathbf{p}$, and the occupied molecular orbital (MO) coefficients of the Slater determinant (SD) denoted by the tensor $\mathbf{C}$, with elements $C_{\mu i}^\sigma$. The indices $\mu, \nu, \lambda$ and $\delta$ label atomic orbitals (AOs), the indices $i, j, k$ and $l$ label occupied MOs, and $\sigma, \tau$ label the spin. Sums over these indices automatically imply complete sums over the total number of occupied orbitals $N_\text{MO}$, atomic orbitals $N_\text{AO}$, and the number of spin channels $N_\text{spin}$, respectively. The overlap matrix between AOs or MOs (depending on its subscripts) is denoted by $S$. In the following, we always label the atomic and molecular orbitals as subscripts, and the spin index as a superscript, where possible.

\vspace{2mm}
Let us start by introducing the VMC Lagrangian:
\begin{eqnarray}\label{eq:lagrangian}
L_{\rm VMC}({\bf p, C, Z, W}) = E_{\rm VMC}(\mathbf{p}, \mathbf{C})
+ \sum_{\lambda k \tau} Z_{\lambda k}^\tau \frac{\partial L_{\rm SCF}(\mathbf{C})}{\partial{C_{\lambda k}^\tau}}
- \sum_{kl\tau} W_{kl}^\tau (S_{kl}^\tau - \delta_{kl}) \; .
\end{eqnarray}
The first term is the VMC energy, the second term corresponds to the stationary conditions of the underlying SCF reference calculation, and the third term enforces the orthonormality of the occupied MOs. The tensors ${\bf Z}$ and ${\bf W}$ contain the Lagrange multipliers. We do not give the explicit form of $L_{\rm SCF}$ here because only its derivative (defined later) is needed. In the most general case, all boldfaced quantities represent tensors of rank 3, with dimension $N_\text{AO}\times N_\text{MO}\times N_\text{spin}$ for $\mathbf{C}$ and $\mathbf{Z}$, and dimension $N_\text{MO}\times N_\text{MO}\times N_\text{spin}$ for $\mathbf{W}$. In case of restricted calculations, where the MO coefficients are the same for both spin channels, $N_\text{spin} = 1$ and the tensors reduce to matrices.
Several equivalent formulations of the SCF term of the Lagrangian exist. In this work, we closely follow the atomic orbital formulation used for the implementation of density-corrected DFT gradients in CP2K~{\cite{2023BEL}}. This results in the Roothaan-Hall equations for the occupied MOs:
\begin{equation}
\frac{\partial L_{\rm SCF}}{\partial C_{\lambda k}^\tau} =
\sum_{\nu} F_{\lambda\nu}^\tau C_{\nu k}^\tau - S_{\lambda\nu} C_{\nu k}^\tau \varepsilon_k^\tau,
\label{eq:roothaan}
\end{equation}
where $F_{\lambda\nu}^\tau$ is the Fock/Kohn-Sham matrix and $\varepsilon_k^\tau$ is the Kohn-Sham eigenvalue of $k$-th MO with the spin $\tau$. The explicit form of the Fock/Kohn-Sham matrix is introduced later.
When the Lagrangian is stationary with respect to all its parameters, we can rewrite $\mathbf{F}^\text{VMC}_\alpha$ with $L_{\rm VMC}$ as:
\begin{eqnarray}\label{eq:lagrangian-force}
\mathbf{F}^{\rm{VMC}}_\alpha = -\cfrac{ {\rm d} E_{\rm VMC} }{ {\rm d}R_\alpha} =
-\cfrac{ {\rm d}L_{\rm VMC} } {{\rm d} R_\alpha} =
-\cfrac{\partial L_{\rm VMC}}{\partial R_\alpha} \; .
\end{eqnarray}
To make the Lagrangian stationary with respect to all its parameters, we have to ensure that the following set of equations is satisfied:
\begin{subequations}
\begin{align}
\frac{\partial L_{\rm VMC}}{\partial p_i} &= \frac{\partial E_{\rm VMC}}{\partial p_i} = 0 \; , \label{eq:vmc_var}
\\
\frac{\partial L_{\rm VMC}}{\partial C_{\mu i}^\sigma} &= \frac{\partial E_{\rm VMC}}{\partial C_{\mu i}^\sigma}
+ \sum_{\lambda k \tau} Z_{\lambda k}^\tau \frac{\partial^2 L_{\rm SCF}}{\partial{C_{\mu i}^\sigma} \partial{C_{\lambda k}^\tau}}
- \sum_{kl\tau} W_{kl}^\tau \frac{\partial S_{kl}^\tau}{\partial C_{\mu i}^\sigma} = 0 \; , \label{eq:mo_var}
\\
\frac{\partial L_{\rm VMC}}{\partial Z_{\lambda k}^\tau} &= \sum_{\nu}
F_{\lambda\nu}^\tau C_{\nu k}^\tau - S_{\lambda\nu} C_{\nu k}^\tau \varepsilon_k^\tau = 0 \; , \label{eq:scf_cond}
\\
\frac{\partial L_{\rm VMC}}{\partial W_{kl}^\tau} &= S_{kl}^\tau - \delta_{kl} = 0 \; . \label{eq:mo_ortho}
\end{align}    
\end{subequations}
Eq.~{\ref{eq:vmc_var}} is the variational condition on the VMC energy with respect to the Jastrow parameters. For an optimized Jastrow wave function, this condition is automatically fulfilled. 
Eq.~{\ref{eq:mo_var}} is zero by virtue of the Lagrange multipliers; that is, we {\it choose} the elements $Z_{\lambda k}^\tau$ and $W_{kl}^\tau$ such that $\tfrac{\partial L_{\rm VMC}}{\partial C_{\mu i}^\sigma} = 0$ for $\forall \mu, i, \sigma$.
Eqs.~{\ref{eq:scf_cond}} and \ref{eq:mo_ortho} are automatically fulfilled if we use canonical orbitals from the underlying SCF calculation.

\vspace{2mm}
At first sight, it seems that Eq.~{\ref{eq:mo_var}} is not sufficient to determine both $\mathbf{Z}$ and $\mathbf{W}$.
However, only certain subblocks are different from zero, such that by projecting the stationary conditions onto occupied and virtual orbital spaces, we are able to obtain enough equations for their determination.
To this end, let us introduce the projection operator $\mathbf{Q}$, with elements:
\begin{equation}\label{eq:Q_projector}
Q_{\mu\nu}^\sigma = \delta_{\mu\nu} - \sum_{\lambda k} C_{\mu k}^\sigma C_{\lambda k}^\sigma S_{\lambda\nu} \; .
\end{equation}
The Lagrange multipliers $Z_{\lambda k}^\tau$ can be best understood as expansion coefficients of the orbital response due to the correlation energy introduced by the Jastrow factor. Hence, only the occupied-virtual blocks contribute to the response, while the occupied-occupied block is assumed to be zero:
\begin{equation}\label{eq:Zij_eq_0}
Z_{ij}^\sigma = \sum_{\mu\nu} Z_{\mu i}^\sigma S_{\mu\nu} C_{\nu j}^\sigma = 0 \; ,
\end{equation}
as unitary rotations among occupied orbitals leave the energy unchanged.
It is convenient at this stage to introduce the explicit form of the Fock/Kohn-Sham matrix:
\begin{equation}\label{eq:Fock_ao}
F_{\mu\nu}^\sigma[\mathbf{P}] = h_{\mu\nu} + \sum_{\lambda\delta\tau} P_{\lambda\delta}^\tau
[(\mu\nu|\lambda\delta) - c_x\delta_{\sigma\tau}(\mu\lambda|\nu\delta)] + V_{\mu\nu}^{\text{xc},\sigma} \; ,
\end{equation}
where $(\mu \nu | \lambda \delta)$ are two-electron integrals over primitive atomic orbitals
\begin{equation}
(\mu\nu|\lambda\delta) = \int\int \psi_{\mu}(\mathbf{r})\psi_{\nu}(\mathbf{r})
\frac{1}{|\mathbf{r} - \mathbf{r}'|} \psi_{\lambda}(\mathbf{r}')\psi_{\delta}(\mathbf{r}')
\mathrm{d}\mathbf{r} \mathrm{d}\mathbf{r}' \; ;
\end{equation}
and $h_{\mu\nu}$ are the kinetic energy and nuclear electron attraction integrals.
This form is completely general and accommodates Hartree-Fock, pure semilocal functionals, and hybrid functionals using the exact exchange parameter $c_x$ and the exchange-correlation potential $V^{\text{xc},\sigma}_{\mu\nu}$.
The tensor $\mathbf{P}$ contains the spin-density matrices in AO basis, with elements:
\begin{equation}\label{eq:density_matrix}
P_{\mu\nu}^\sigma = \sum_{i} C_{\mu i}^\sigma C_{\nu i}^\sigma \; .
\end{equation}
The stationary conditions for the MO coefficients are given by equating to zero the derivatives of $L_{\rm VMC}$ with respect to $C_{\mu i}^\sigma$:
\begin{eqnarray}\label{eq:lagrangian-multiplier}
\frac{\partial L_{\rm VMC}}{\partial C_{\mu i}^\sigma} = \frac{\partial E_{\rm VMC}}{\partial C_{\mu i}^\sigma}
+ \sum_{\lambda k \tau} Z_{\lambda k}^\tau \sum_{\nu} \frac{\partial}{\partial C_{\mu i}^\sigma} 
(F_{\lambda\nu}^\tau C_{\nu k}^\tau - S_{\lambda\nu} C_{\nu k}^\tau \varepsilon_k^\tau)
- \sum_{kl\tau} W_{kl}^\tau \frac{\partial S_{kl}^\tau}{\partial C_{\mu i}^\sigma} = 0 \; .
\end{eqnarray}
To arrive at working equations to determine the multipliers, we project Eq.~{\ref{eq:lagrangian-multiplier}} onto the occupied orbital space using $\mathbf{C}$, and onto the unoccupied orbital space using $\mathbf{Q}$:
\begin{subequations}
\begin{align}
\sum_{\mu} \frac{\partial L_{\rm VMC}}{\partial C_{\mu i}^\sigma} Q_{\mu \nu}^\sigma &= 0 \to Z_{\nu i}^\sigma \quad \forall \, \nu, i, \sigma \; ,
\label{eq:L_VMC_onto_virt}
\\
\sum_{\mu} \frac{\partial L_{\rm VMC}}{\partial C_{\mu i}^\sigma} C_{\mu j}^\sigma &= 0 \to W_{ij}^\sigma \quad \forall \, i, j, \sigma \; .
\label{eq:L_VMC_onto_occ}
\end{align}
\end{subequations}
In the following, we report only the final expressions obtained from Eqs.~\ref{eq:L_VMC_onto_virt} and \ref{eq:L_VMC_onto_occ}; their explicit derivation is given in Appendix~\ref{app:lagrange}.
The projection onto the virtual space, Eq.~\ref{eq:L_VMC_onto_virt}, results in a linear system of equations $\mathbf{A} \mathbf{Z} = -\mathbf{B}$ known as the Z-vector equations, from which the Lagrange multipliers $\mathbf{Z}$ can be determined:
\begin{equation}\label{eq:AZ_eq_B_elements}
\sum_\lambda Z_{\lambda i}^\sigma (F_{\lambda\mu}^\sigma - S_{\lambda\mu} \varepsilon_i^\sigma)
+ \sum_{\mu} H_{\mu i}^\sigma \big[ \bar{\mathbf{P}} \big] Q_{\mu\nu}^\sigma = - \sum_{\mu} B_{\mu i}^\sigma Q_{\mu\nu}^\sigma \; .
\end{equation}
The elements of the tensor $\mathbf{B}$ are the derivatives of the VMC energy with respect to the MO coefficients
\begin{equation}
B_{\mu i}^\sigma = \frac{\partial E_{\rm VMC}}{\partial C_{\mu i}^\sigma} \; ,
\end{equation}
while the tensor $\mathbf{A}$ contains quantities related to the SCF stationary conditions and the MOs orthonormalization constraints; hence it does not depend on the VMC calculation.
The elements $H_{\mu i}^\sigma \big[ \bar{\mathbf{P}} \big]$ are derivatives of the Fock matrix with respect to the MO coefficients,
\begin{equation}\label{eq:H_mu_i}
H_{\mu i}^\sigma \big[ \bar{\mathbf{P}} \big] = \sum_{\lambda\nu\tau} \bar{P}_{\lambda\nu}^\tau
(2(\lambda\nu | \mu i_\sigma ) - c_x\delta_{\tau\sigma} [(\lambda \mu | \nu i_\sigma) + (\lambda i_\sigma| \nu \mu)])
+ 2f_{\lambda\nu\tau, \mu i \sigma}^{\text{xc}} \; ;
\end{equation}
where $f^\text{xc}_{\lambda\nu\tau,\mu i\sigma}$ is the exchange-correlation kernel with an index contracted with a molecular orbital, and the tensor $\bar{\mathbf{P}}$ is a symmetrized form of the response density matrix (for increased numerical stability), given by
\begin{equation}\label{eq:P_bar}
\bar{P}_{\mu\nu}^\sigma = \frac{1}{2} \sum_i (Z_{\mu i}^\sigma C_{\nu i}^\sigma + C_{\mu i}^\sigma Z_{\nu i}^\sigma ) \; .
\end{equation}
Once the response tensor $\mathbf{Z}$ is determined by solving Eq.~\ref{eq:AZ_eq_B_elements}, we can also determine the second set of Lagrange multipliers $\mathbf{W}$, which are given by
\begin{equation}\label{eq:Wij}
W_{ij}^\sigma = \frac{1}{2} H_{ij}^\sigma \big[ \bar{\mathbf{P}} \big]
+ \frac{1}{2}\sum_{\mu} B_{\mu i}^\sigma C_{\mu j}^\sigma.
\end{equation}
Finally, the atomic force is given by the negative gradient of the Lagrangian with respect to the nuclear positions, reading
\begin{eqnarray}\label{eq:lagrangian_derivative}
\frac{\partial L_{\rm VMC}}{\partial \mathbf{R}_\alpha} =
\frac{\partial E_{\rm VMC}}{\partial \mathbf{R}_\alpha} -
\sum_{\mu \nu \sigma} \Lambda_{\mu\nu}^\sigma \frac{\partial S_{\mu\nu}}{\partial \mathbf{R}_\alpha} +
\sum_{\mu \nu j \sigma} Z_{\mu j}^\sigma  \frac{\partial F_{\mu\nu}^\sigma}{\partial \mathbf{R}_\alpha} C_{\nu j}^\sigma
\; ,
\end{eqnarray}
with the intermediate quantity $\Lambda_{\mu \nu}^\sigma$ defined as
\begin{equation}
\Lambda_{\mu\nu}^\sigma = 
\sum_{ij} C_{\mu i}^\sigma W_{ij}^\sigma C_{\nu j}^\sigma + \frac{1}{2}
\sum_i \varepsilon_{i}^\sigma (Z_{\mu i}^\sigma C_{\nu i}^\sigma + C_{\mu i}^\sigma Z_{\nu i}^\sigma) \; ,
\end{equation}
embodying the conditions of orthonormality between the response MOs, and the so-called energy-weighted response density matrix.

\vspace{2mm}
An important aspect of the above Lagrangian approach in the context of VMC is that the elements of the right-hand side tensor $\mathbf{B}$ of Eq.~\ref{eq:AZ_eq_B_elements} (i.e., the energy derivative with respect to the MO coefficients) come with statistical noise. It is important to know how this noise propagates from the VMC energy derivative to the response vectors, and ultimately to the forces. There are several ways to estimate the errors: the easiest way is to use a resampling method such as the Jackknife approach. As an outcome of a VMC calculation, we can reblock samples of $B^{\sigma}_{\mu i}$ and estimate the error bars of the corrections by iteratively solving the Z-vector equations. Another way to estimate the errors is based on the analytical error propagation of $B^{\sigma}_{\mu i}$, for which the variance-covariance tensor of $B^{\sigma}_{\mu i}$ is required.
In this paper, we employed the former approach, i.e., the error bar estimations were done using the Jackknife method.
Further details about the VMC energy and gradients calculations, as well as the derivation of the analytical error propagation is shown in Appendix~{\ref{app:vmc}}.

\section{Validation of the method}
\label{sec:validation}
In this section, we validate the Lagrangian technique described above, which we have implemented by interfacing the linear response module of CP2K~\cite{Kuehne2020_CP2K_JCP} with TurboRVB~\cite{2020NAK2} and exchanging the data through the TREX-IO library~\cite{2023POS}. The validation means demonstrating two points: (i) forces and pressures obtained with the Lagrangian technique agree with numerical derivatives of the potential energy surface (i.e., the results are unbiased), and (ii) these forces and pressures are consistent with those computed by the finite-difference method (FDM) developed in Ref.~\onlinecite{Nakano2024}. The validation was performed for a Cl$_2$ molecule, crystalline c-BN, and acetamide (CH$_3$CONH$_2$ $\equiv$ AcNH$_2$) dimer. For Cl$_2$, we evaluated forces; for c-BN, we evaluated both forces and pressure; for AcNH$_2$ dimer, we evaluated forces along the O--H bonds acting on each fragment and total forces. For Cl$_2$, we used the ccECP pseudopotentials~{\cite{2018BEN}} together with their accompanying cc-pVDZ basis sets. For c-BN, we used the ccECP~{\cite{2017BEN}} pseudopotentials combined with DZVP-MOLOPT-SR-GTH basis sets~{\cite{Goedecker1996_GTH, VandeVondele2007_MOLOPT}} to avoid spuriously large force error bars by GTOs with large exponents~{\cite{Nakano2021PRB_finite_variance_of_forces_for_PBC}}. The cBN calculations employed a conventional 1 $\times$ 1 $\times$ 1 supercell with $\Gamma$-point.
For AcNH$_2$ dimer, we used the ccECP pseudopotentials~{\cite{2018BEN}} together with their accompanying cc-pVDZ basis sets. To estimate the statistical error of the linear-response (LR) correction, we used the jackknife approach~{\cite{2017BEC}}: we repeatedly applied the LR procedure to samples of $\partial E / \partial P_{\mu\nu}^{\uparrow\downarrow}$ taken from TurboRVB and computed the jackknife mean and variance.
Figures~{\ref{fig:Cl2_fdm_vs_lr}}, {\ref{fig:cBN_fdm_vs_lr}} and \ref{fig:AcNH2_lr} summarize the results for the Cl$_2$ forces, for the c-BN forces and pressure, and for AcNH$_2$ dimer forces, addressing points (i) and (ii). As shown, the forces and pressures obtained by the Lagrangian technique are consistent with the numerical derivatives of the potential energy surface and match the values obtained with the FDM~{\cite{Nakano2024}}. These results confirm that the Lagrangian technique works as expected. The inset of
fig.~{\ref{fig:AcNH2_lr}} also demonstrates that the total forces obtained by the Lagrangian technique are always zero, as they should be.
Notice that we do not compare the obtained PESs and EOS with the experimental bond lengths and volume, since the used basis sets, the supercell size, and the $k$-point ($\Gamma$ only) are suitable only for numerically checking (i) and (ii).

\begin{figure}
    \centering
    \includegraphics[width=1.0\linewidth]{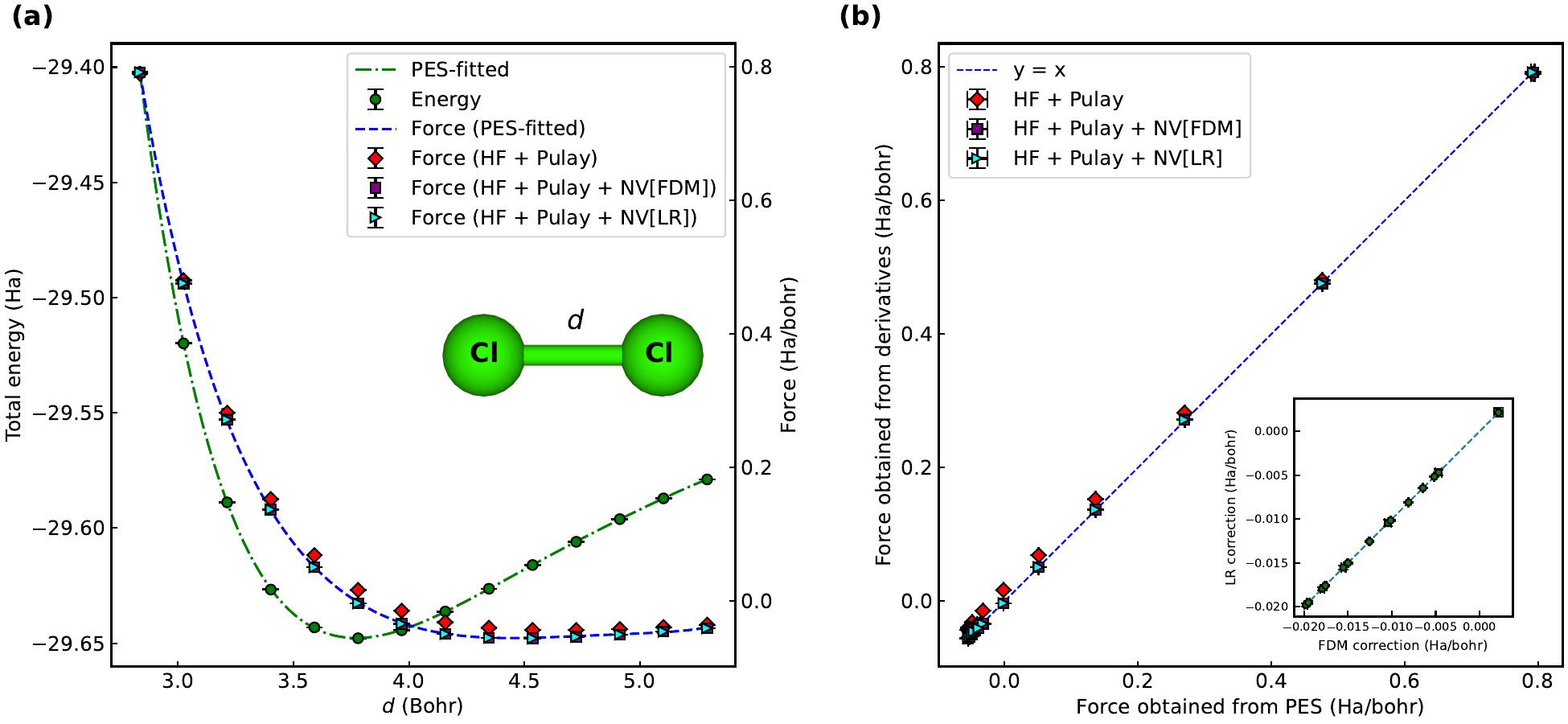}
    \caption[]{Validation of the Lagrangian technique for Cl$_2$. (a) Potential energy surface (PES) as a function of bond length (green circles) and its fit (green dotted line). The reference force on the right-hand Cl atom (Cl$_2$ oriented along the $x$ axis), obtained from the derivative of the fitted PES, is shown as a blue dashed line. Red diamonds, purple squares, and cyan triangles denote forces evaluated with the HF + Pulay terms, HF + Pulay terms + NV correction (via FDM), and HF + Pulay terms + NV correction (via LR), respectively. (b) Comparison between the PES-derived (reference) forces and those obtained from the estimators. Inset: direct comparison of the NV corrections via FDM and LR.
    }
    \label{fig:Cl2_fdm_vs_lr}
\end{figure}

\begin{figure}
    \centering
    \includegraphics[width=1.0\linewidth]{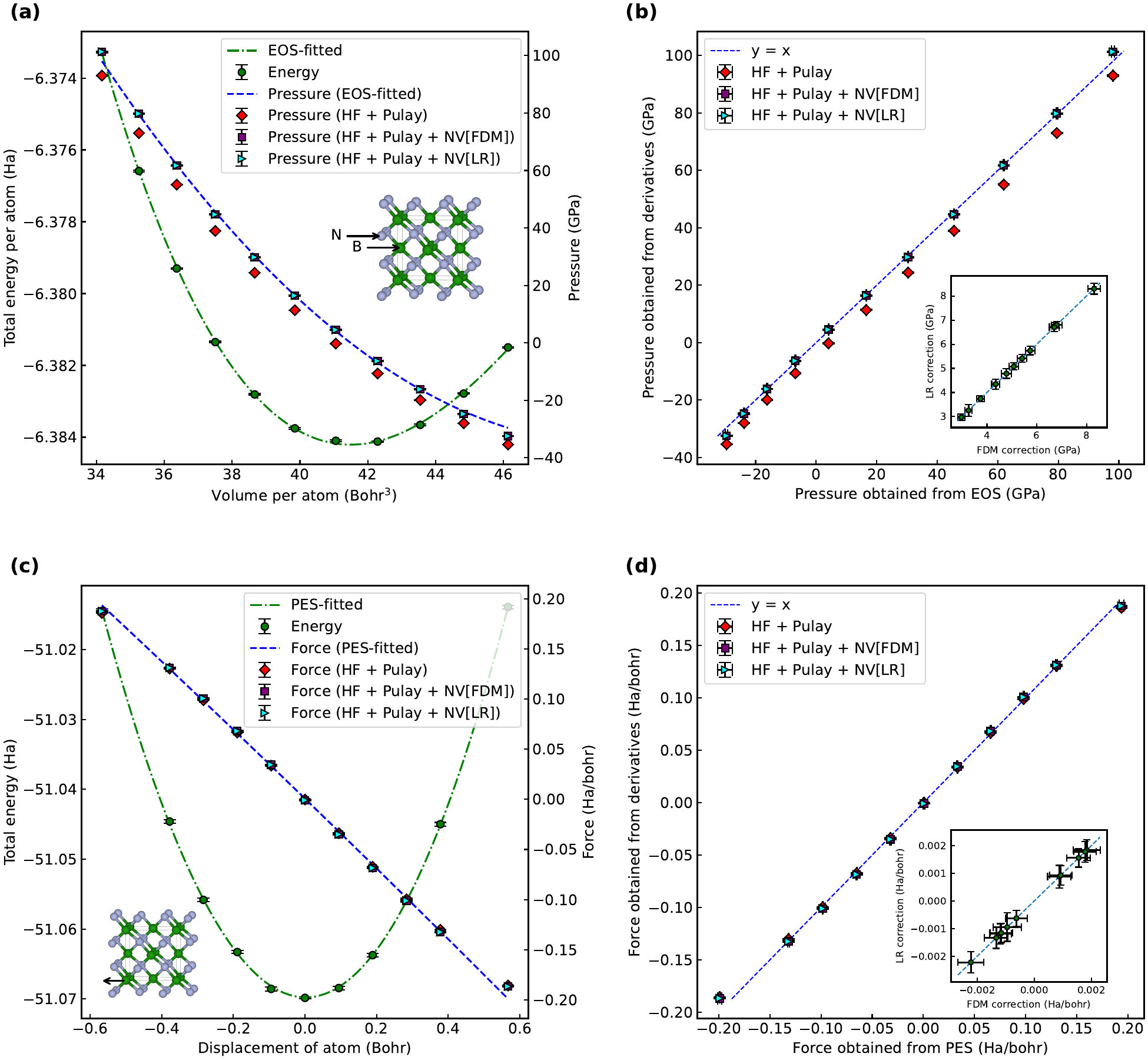}
    \caption[]{Validation of the Lagrangian technique for cBN. (a) Equation of state (EOS) as a function of volume (green circles) and its fit (green dotted line). The reference pressure obtained from the derivative of the fitted EOS, is shown as a blue dashed line. Red diamonds, purple squares, and cyan triangles denote pressures evaluated with the HF + Pulay terms, HF + Pulay terms + NV correction (via FDM), and HF + Pulay terms + NV correction (via LR), respectively. (b) Comparison between the EOS-derived (reference) pressures and those obtained from the estimators. Inset: direct comparison of the NV corrections via FDM and LR.
    (c) PES as a function of a displacement of the B atom located at the origin (0,0,0) of the unit cell in the $x$ direction (green circles) and its fit (green dotted line). The reference force on the B atom, obtained from the derivative of the fitted PES, is shown as a blue dashed line. Red diamonds, purple squares, and cyan triangles denote forces evaluated with the HF + Pulay terms, HF + Pulay terms + NV correction (via FDM), and HF + Pulay terms + NV correction (via LR), respectively. (d) Comparison between the PES-derived (reference) forces and those obtained from the estimators. Inset: direct comparison of the NV corrections via FDM and LR.
    }
    \label{fig:cBN_fdm_vs_lr}
\end{figure}

\begin{figure}
    \centering
    \includegraphics[width=1.0\linewidth]{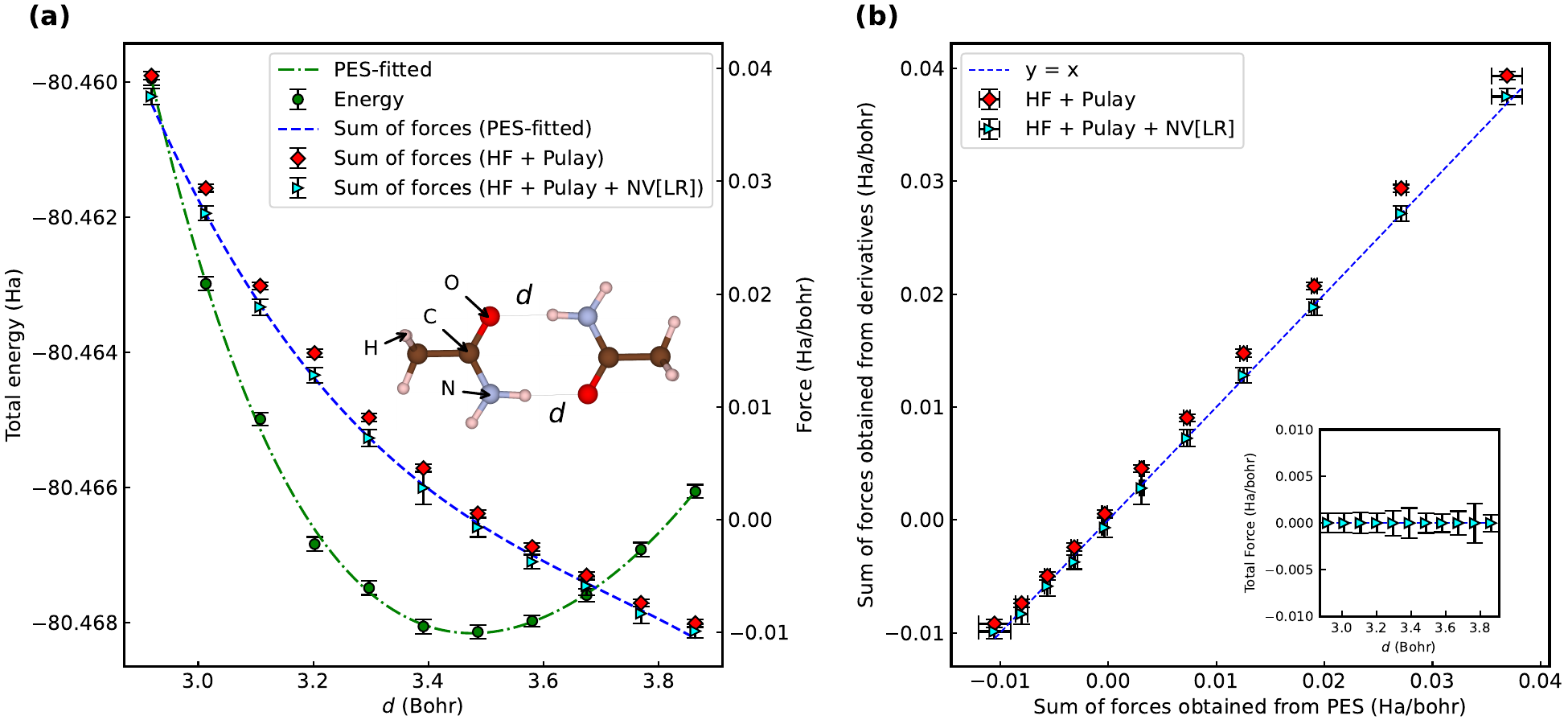}
    \caption[]{Validation of the Lagrangian technique for AcNH$_2$ dimer. (a) Potential energy surface (PES) as a function of O--H bond length (green circles) denoted as $d$ and its fit (green dotted line). The reference force (i.e. the sum of forces acting on the right fragment along the two O--H bonds, where the two O--H bonds are aligned along the $x$ axis), obtained from the derivative of the fitted PES, is shown as a blue dashed line. Red diamonds and cyan triangles denote the sum of forces in the $x$ direction evaluated with the HF + Pulay terms and HF + Pulay terms + NV correction (via LR), respectively. (b) Comparison between the PES-derived (reference) forces and those obtained from the estimator, HF + Pulay terms + NV correction (via LR). Inset:    
    The total forces of the system along the O--H bonds (i.e., in the $x$ direction), which should be zero, as a function of O--H bond length.
    }
    \label{fig:AcNH2_lr}
\end{figure}

\vspace{2mm}
We next discuss the computational cost of the Lagrangian technique. The primary advantage of the Lagrangian technique is efficiency: unlike the FDM approach~{\cite{Nakano2024}}, which requires 6$N$ separate DFT calculations to obtain the forces, the Lagrangian technique needs only a single-shot CPHF/CPKS calculation. On the one hand, the FDM approach results in an $O(N^4)$ asymptotic scaling because it requires 6$N$ times DFT calculations and the computational complexity of DFT is typically $O(N^3)$. On the other hand, a CPHF/CPKS calculation shows the same asymptotic scaling as a single-shot HF/KS-SCF calculation does, implying that the asymptotic scaling of the Lagrangian technique remains $O(N^3)$. Therefore, the Lagrangian approach is more affordable as the system size increases. We note that the dominant computational cost for atomic force calculations in total arises from the VMC step, not from the LR calculation step. Therefore, the feasibility of unbiased VMC force calculations is determined by the cost of the corresponding biased VMC force calculations

\vspace{2mm}
We did a benchmark test for the LR calculation using CP2K for cBN systems with $N$ = $k$ $\times$ $k$ $\times$ $k$ $\times$ 8 atoms in a simulation cell. We measured the timing of DFT and LR calculations on a dual-socket AMD EPYC 7742 64-Core Processor for $k$ from 1 to 8. The DZVP basis (13 bsf/atom) is employed with the ccECP pseudopotential. Table~{\ref{tab:comparison_dft_lr}} summarizes the timings of the calculations, and Figure~{\ref{fig:dft_vs_lr}} plots the scaling analysis. The scaling analysis reveals that, in the region we tested, DFT and LR exhibit $N^{\sim 3}$ and $N^{\sim 2}$ behavior, respectively, which are $N$ order of magnitude smaller than the expected asymptotic scalings. This is probably because the investigated $N$ are too small to see the expected asymptotic scaling. However, it demonstrates that the prefactor and scaling of a single-shot LR are better than those of 6$N$-times DFT calculations. The systematic test shows that our method is faster than the FDM method relying on the 6$N$-times DFT calculations.
%

\begin{table}
  \caption{The comparison of timings of 6$N$-times DFT calculations and single-shot LR calculation for cBN systems with $N$ = $k$ $\times$ $k$ $\times$ $k$ $\times$ 8 atoms. The DZVP basis (13 bsf/atom) is employed with ccECP pseudo potetnial. The timings were measured in minutes on dual socket AMD EPYC 7742 64-Core Processor.}
  \label{tab:comparison_dft_lr}
  \centering
  \begin{tabular}{c|c|cc}
    \Hline
    $k$ & $N$ & 6$N$-times DFT & Single-shot LR \\
    \Hline
    1 & 8 & 4.99 & 0.13 \\
    2 & 64 & 47.60 & 0.21 \\
    3 & 216 & 441.33 & 0.57 \\
    4 & 512 & 3130.68 & 1.70 \\
    5 & 1000 & 21095.40 & 5.39 \\
    6 & 1728 & 94109.82 & 14.11 \\
    7 & 2744 & 392623.87 & 37.12 \\
    8 & 4096 & 1628106.34 & 133.98 \\
    \Hline
  \end{tabular}
\end{table}

\begin{figure}
    \centering
    \includegraphics[width=0.5\linewidth]{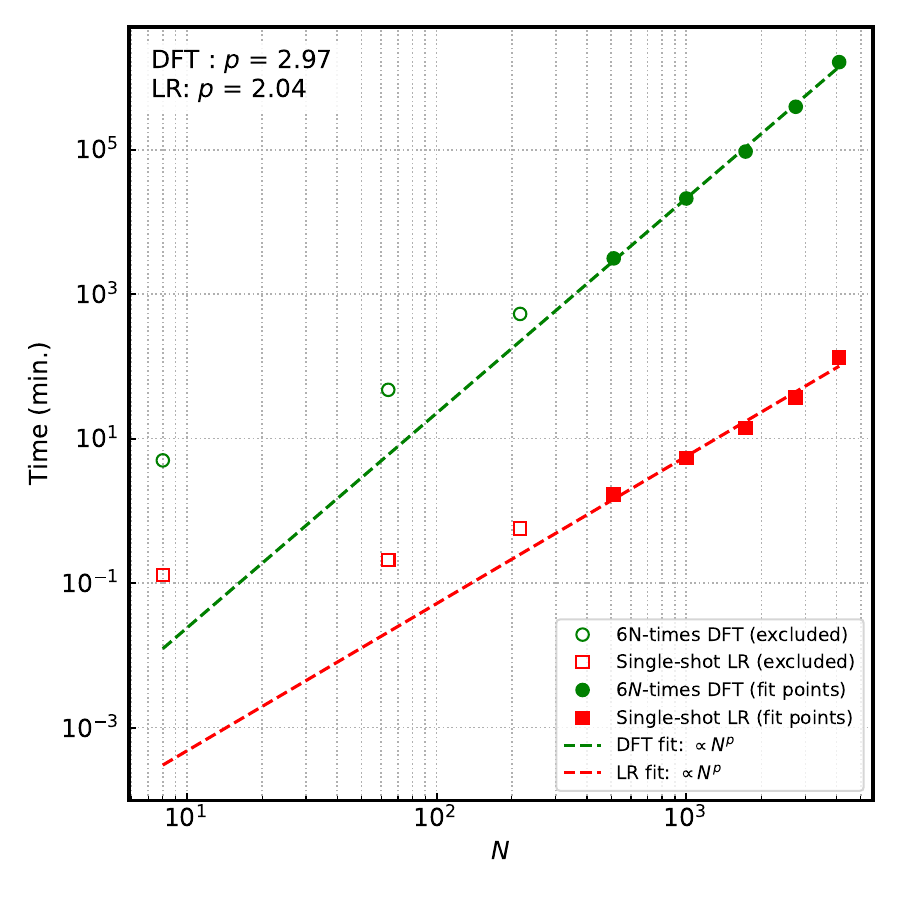}
    \caption[]{The scaling analysis of DFT and LR calculations with respect to the number of atoms ($N$) in the cBN simulation cell.}
    \label{fig:dft_vs_lr}
\end{figure}

\vspace{2mm}
From a technical point of view, the exchange of wavefunctions between CP2K~{\cite{Kuehne2020_CP2K_JCP}} and TurboRVB~{\cite{2020NAK2}} required careful consideration to deal with differences in conventions, which is a common challenge among electronic structure packages. Examples are the use of spherical-harmonic versus Cartesian Gaussian Type Orbitals (GTOs), normalization constants, and their application stage (pre- vs post-contraction, shell-wise vs AO-level), and related notational choices. In this work, we adopted the TREX-IO~{\cite{2023POS}} format, a flexible library that standardizes conventions for exchanging wavefunctions, structures, and pseudopotentials among electronic structure packages. TREX-IO effectively absorbed inter-code convention mismatches and played a key role in robust data exchange between CP2k and TurboRVB in this work.

\section{Application to organic molecules in rMD17 dataset}
\label{sec:application}
Now that we have established an unbiased VMC force evaluation based on the Lagrangian technique, a systematic assessment of VMC force accuracy is feasible across a wide range of compounds. VMC and DMC results are increasingly being used as training dataset for MLPs~{\cite{2022TIR_MLP_hydrogen, 2023NIU_qmc_hydrogen, 2024CEP_qmc_hydrogen, 2024GIA_qmc_hugonioit, slootman2024accurate, 2025Giacomo, Goswami2025_hydrogen_MLP, Kevin2025_hydrogen_RQMC}}. In this study, we selected ethanol, malonaldehyde, and benzene from the rMD17 dataset~{\cite{Christensen2020_MLST, Christensen2020_rMD17_Figshare}}, and computed (i) unbiased VMC forces established in this work, (ii) VMC forces with fully optimized wavefunctions (iii) Hartree-Fock (HF), second-order M{\o}ller--Plesset perturbation theory (MP2), CCSD, and DFT forces, and (iv) CCSD(T) forces as references.
We took 100 structures from molecular dynamics trajectories distributed as rMD17 dataset~{\cite{Christensen2020_MLST, Christensen2020_rMD17_Figshare}}. In a previous study that constructed MLPs with CCSD(T) references~{\cite{2018Stefan}}, the Dunning cc-pVTZ~{\cite{1989DUN}} was used for ethanol, and cc-pVDZ~{\cite{1989DUN}} for malonaldehyde and benzene to generate a training dataset for MLPs. Here, we used cc-pVQZ for the three molecules for the CCSD(T) calculations to obtain forces closer to the complete basis-set limit. For comparison, we also carried out HF, MP2~{\cite{1934MOL_MP2}}, and CCSD~{\cite{Purvis1982_CCSD}} with the cc-pVQZ basis and DFT calculations with the def2-QZVPPD~{\cite{Weigend2005_def2}} basis set using several exchange–correlation functionals with the D3BJ dispersion correction~{\cite{Grimme2010_D3, Grimme2011_D3BJ}}. Specifically, we chose SVWN~{\cite{Vosko1980_VWN}}, PBE~{\cite{1996PER}}, PBE+D3BJ, PBE0~{\cite{Adamo1999_PBE0}}+D3BJ, $\omega$B97X+D3BJ~{\cite{Mardirossian2014_wB97X_V, Najibi2018_wB97X_wB97M_D3BJ}}, and $\omega$B97M+D3BJ~{\cite{Mardirossian2016_wB97M_V, Najibi2018_wB97X_wB97M_D3BJ}}. In the case of MP2, CCSD, and CCSD(T) calculations, the 1$s$ shells were kept frozen.
All of these calculations were performed with Psi4 software suite (Version 1.9.1)~{\cite{Smith2020_PSI4_1_4}}.
We notice that, for each of ethanol, malonaldehyde, and benzene, the first structure was also computed with Molpro (Version~2025.3)~{\cite{MOLPRO-WIREs, MOLPRO-JCP, MOLPRO}} and PySCF (Version~2.10.0)~{\cite{2018SUN_pyscf, 2020SUN_pyscf}} to double-check the Psi4 results. We confirmed that Psi4, PySCF, and Molpro yielded consistent energies and forces for all three molecules.
SVWN and PBE were selected as computationally inexpensive exchange–correlation functionals. PBE0 was included to allow comparison with a previous VMC/DMC benchmark study~{\cite{slootman2024accurate}}, which combined Tkatchenko-Scheffler (TS)~{\cite{2009ALEX-TS}} and the many-body dispersion (MBD)~{\cite{2012ALEX-MBD}} corrections with PBE and PBE0, respectively. Since the Psi4 software suite does not implement TS or MBD, we instead employed D3BJ dispersion correction in this work. The $\omega$B97 family of functionals was chosen because it is often used in popular MLP datasets (e.g., SPICE~{\cite{Eastman2023_SPICE_SciData}}).
For the VMC calculations, we employed ccECP pseudopotentials~{\cite{2017BEN}} together with the accompanied cc-pVTZ basis set. Following the previous VMC/DMC benchmark study by Slootman et al.~{\cite{slootman2024accurate}}, we also benchmarked ccECP-based CCSD against all-electron CCSD using PySCF~{\cite{2018SUN_pyscf, 2020SUN_pyscf}} and confirmed that the force errors for the three molecules are $<$ 1 kcal/mol/\AA\ in root-mean-square errors (RMSEs). In the VMC calculations for the unbiased forces, the Slater determinant part was fixed to orbitals obtained from DFT (LDA-PZ), and only the Jastrow part was variationally optimized using stochastic reconfiguration~{\cite{1998SOR}} with an optimal step size. 
In the VMC calculations with fully optimized wavefunction, both the molecular orbital coefficients in the Slater determinant and the Jastrow part were variationally optimized.
The Jastrow factor comprised a two-body term satisfying the electron–electron cusp~{\cite{2020NAK2}}, and an inhomogeneous one-body term and a three-body term parametrized using localized atomic orbitals~{\cite{2020NAK2}}. For the inhomogeneous one-body and three-body Jastrow basis sets, we used uncontracted primitive GTOs with [3s1p] for H and [4s2p1d] for C, N, and O; the corresponding Gaussian exponents were optimized for each configuration during the optimization procedure. Although matrix elements coupling different atomic pairs (often referred to as four-body Jastrow terms~{\cite{2020NAK2}}) are frequently set to zero in practice~{\cite{2020NAK2}}, we retained them here because they have an impact on the forces (See. Appendix~{\ref{app:rmd17-parameter-study}}). The total numbers of optimized variational parameters were 3422, 5069, and 8140 for ethanol, malonaldehyde, and benzene, respectively. The trial wavefunctions were generated with CP2K~{\cite{Kuehne2020_CP2K_JCP}}; VMC simulations were performed with TurboRVB~{\cite{2020NAK2}}; and the force-bias correction based on the Lagrangian technique was applied via CP2K~{\cite{Kuehne2020_CP2K_JCP}}.

\begin{table}[htbp]
\centering
\caption{The RMSE of atomic forces of all the configurations for the selected methods relative to CCSD(T). The unit is kcal/mol/\AA.}
\label{tab:rmse}
\begin{tabular}{lccc}
\Hline
Method & ethanol & malonaldehyde & benzene \\
\Hline
VMC(biased) & 2.924(9) & 9.56(1) & 3.19(1) \\
VMC(unbiased) & 2.549(9) & 7.00(1) & 2.18(1) \\
VMC(optimized) & 2.455(9) & 7.19(1) & 2.54(1) \\
\hline
HF & 9.45 & 23.44 & 5.57 \\
MP2 & 1.39 & 2.13 & 1.64 \\
CCSD & 1.62 & 5.01 & 2.00 \\
SVWN & 7.37 & 8.86 & 8.32 \\
PBE & 4.63 & 4.72 & 3.09 \\
PBE-D3BJ & 4.60 & 4.68 & 3.06 \\
PBE0-D3BJ & 2.21 & 7.80 & 3.75 \\
$\omega$B97X-D3BJ & 0.98 & 5.40 & 3.13 \\
$\omega$B97M-D3BJ & 1.24 & 6.31 & 3.08 \\
\Hline
\end{tabular}
\end{table}

\begin{figure}
    \centering
    \includegraphics[width=0.9\linewidth]{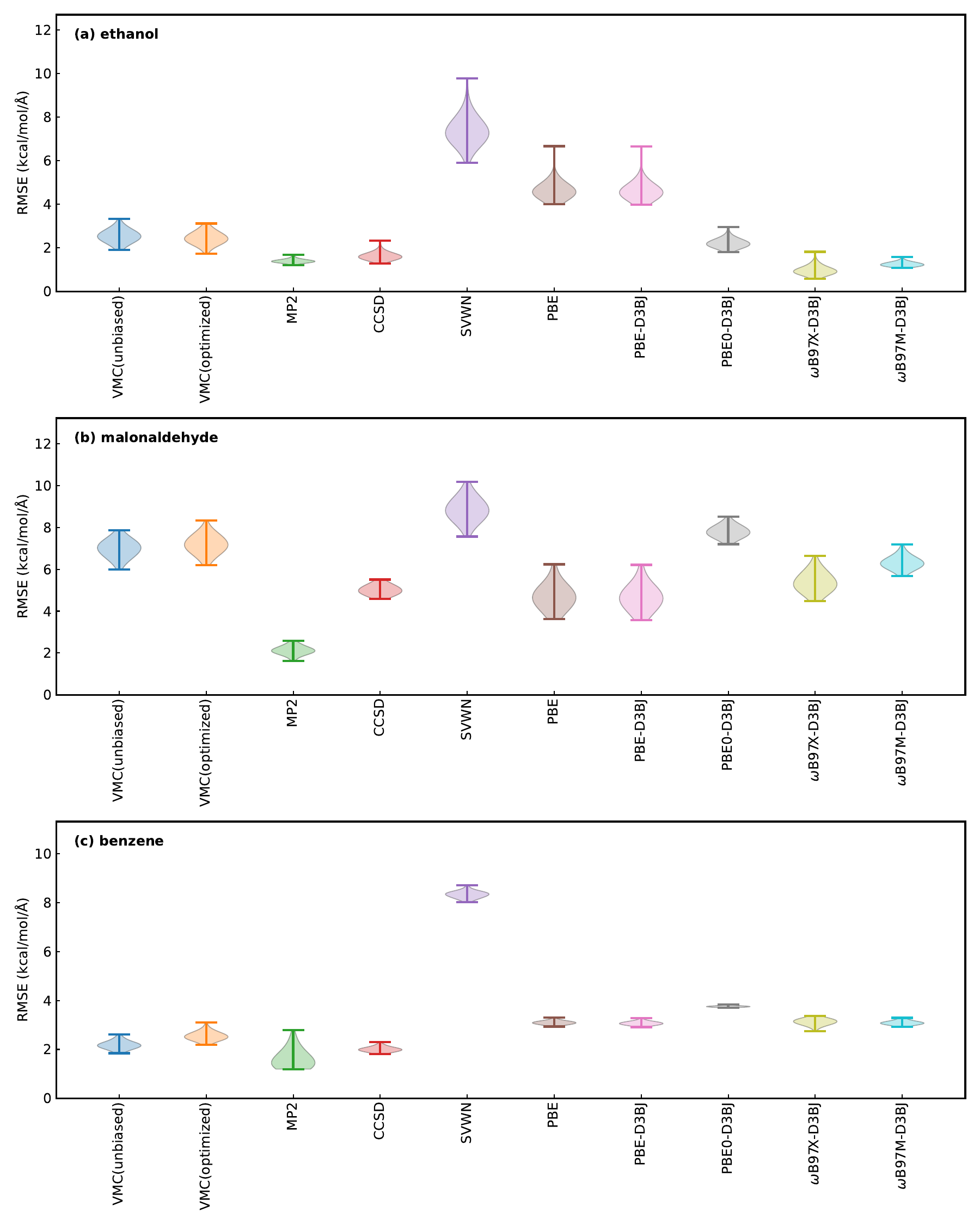}
    \caption[]{Violin plots showing, for each configuration of (a) Ethanol, (b) Malonaldehyde, and (c) Benzene, the RMSE of atomic forces for the selected methods relative to CCSD(T). The unit is kcal/mol/\AA.}
    \label{fig:RMSE}
\end{figure}

{\vspace{2mm}}
Table~{\ref{tab:rmse}} reports the RMSE of all force components (i.e., $N_{\mathrm{atoms}}\times 3$ directions $\times$ 100 structures) relative to CCSD(T) for each molecule. Figure~{\ref{fig:RMSE}} shows, for each molecule, the violin plot of the distribution of RMSE in 100 configurations for selected methods. We first emphasize that the benefit of the Lagrangian technique is evident from Table~{\ref{tab:rmse}}. Comparing the CCSD(T) reference with biased VMC forces (i.e., without the Lagrangian-technique correction), the RMSEs are 2.924(9), 9.56(1), and 3.19(1) kcal mol$^{-1}$ Å$^{-1}$ for ethanol, malonaldehyde, and benzene respectively. In contrast, the unbiased VMC forces developed in this study yield RMSEs of 2.549(9), 7.00(1), and 2.18(1) kcal mol$^{-1}$ \AA$^{-1}$ for ethanol, malonaldehyde, and benzene respectively. They suggest that, when accurate forces are sought with only partially optimized wavefunctions, our correction scheme is essential. 
The VMC forces obtained with the fully optimized wavefunction yield RMSEs of 2.455(9), 7.19(1), and 2.54(1) kcal mol$^{-1}$ \AA$^{-1}$ for ethanol, malonaldehyde, and benzene respectively. They are very close to the unbiased VMC ones, demonstrating that, {\it even without optimizing the determinant part}, correcting the NV term can achieve force accuracies comparable to those obtained with fully optimized wavefunctions, thereby reducing optimization effort.

\vspace{2mm}
For ethanol, Slootman et al.~{\cite{slootman2024accurate}} performed an extensive benchmark for a set of 200 configurations and reported an RMSE of 3.055 kcal mol$^{-1}$ \AA$^{-1}$ for VMC forces based on a fully optimized Jastrow Slater wavefunction. Our ethanol RMSEs, 2.549(9)~kcal mol$^{-1}$ \AA$^{-1}$ with the frozen DFT orbitals and 2.455(9) ~kcal mol$^{-1}$ \AA$^{-1}$ with the fully optimized Jastrow Slater wavefunction, are better than their values. This improvement is probably due to the much larger Jastrow factor used in this study. We note that our ethanol configurations are not identical to theirs; thus, a strict one-to-one comparison is not possible.
If one pursues the force accuracy of $\sim$ 1 kcal mol$^{-1}$ \AA$^{-1}$ relative to CCSD(T), DMC calculations would be required, as revealed by Slootman et al.~{\cite{slootman2024accurate}}. While the present Lagrangian technique is applicable to DMC force calculations in principle, a full DMC implementation is beyond the scope of this paper and is left for future work.

{\vspace{2mm}}
Table~{\ref{tab:rmse}} also lists the HF, MP2, CCSD, and other DFT results for all three molecules. For ethanol, $\omega$B97X-D3BJ and $\omega$B97M-D3BJ yield $\sim$ 1 kcal mol$^{-1}$ \AA$^{-1}$ RMSEs relative to CCSD(T). Slootman et al.~{\cite{slootman2024accurate}} compared VMC with dispersion-corrected PBE-TS and PBE0-MB, finding RMSE(PBE0-MBD) $<$ RMSE(VMC) $<$ RMSE(PBE-TS); that is, VMC performed between a GGA and a hybrid GGA. Our results reproduce this trend.
Interestingly, this trend is not universal across the molecules. 
For benzene, the VMC-unbiased and VMC-optimized RMSEs are 2.18(1) kcal mol$^{-1}$ \AA$^{-1}$ and 2.54(1) kcal mol$^{-1}$ \AA$^{-1}$, respectively, whereas $\omega$B97X-D3BJ and $\omega$B97M-D3BJ are worse than the VMC ones, exceeding 3 kcal mol$^{-1}$ \AA$^{-1}$. 
For malonaldehyde, both the VMC-unbiased and VMC-optimized RMSEs yield large RMSEs, 7.00(1) kcal mol$^{-1}$ \AA$^{-1}$ and 7.19(1) kcal mol$^{-1}$ \AA$^{-1}$, respectively. Not only the VMCs but also PBE0-D3BJ and $\omega$B97 functional family give large RMSEs in order of $\sim$ 5-8 kcal mol$^{-1}$ \AA$^{-1}$.
For all three compounds, HF or SVWN shows the worst RMSE among the tested methods, while MP2 shows a very good performance.
These results demonstrate that our unbiased VMC forces appear advantageous over biased VMC ones in terms of accuracy. However, VMC is outperformed by MP2 for all three compounds. Although VMC offers a more favorable scaling with the number of electrons (typically, $O(N^{3})$~\cite{esler2008quantum}} and $O(N^{4-5})$ for VMC and MP2, respectively) and is naturally suitable for parallelization, the large prefactor in practical VMC implementations complicates any conclusion about superior wall-time efficiency (e.g., VMC is roughly a factor of 100 heavier than MP2 for ethanol).

\begin{figure}
    \centering    \includegraphics[width=1.00\linewidth]{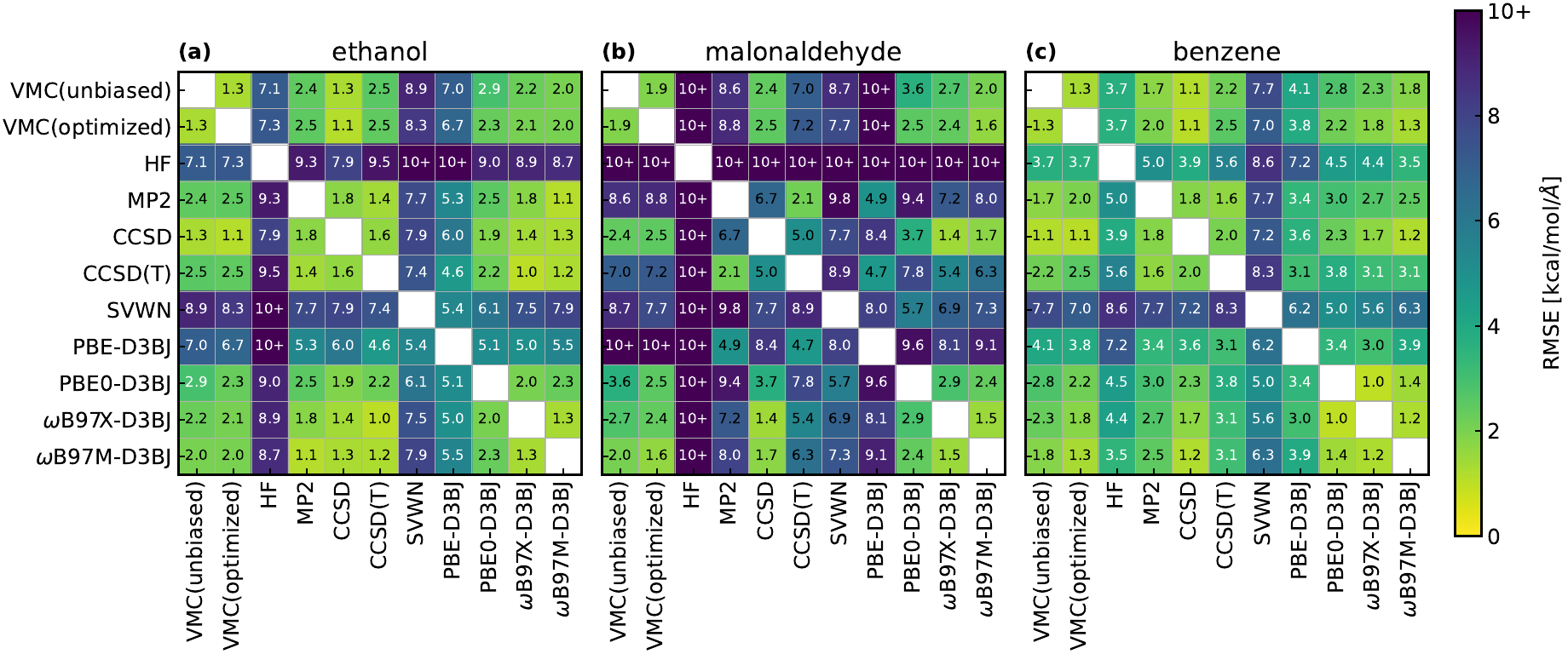}
    \caption[]{Pairwise RMSEs of atomic forces (kcal/mol/\AA) between all pairs of the benchmarked methods for the three molecules: (a) Ethanol, (b) Malonaldehyde, and (c) Benzene. The color scale is shared among the plots and is saturated at 10~kcal/mol/\AA\ (i.e., values $\ge 10$ are mapped to the same color). The numbers indicate the corresponding RMSEs for each pair.}
    \label{fig:pairwise-rmse-heatmap}
\end{figure}

\vspace{2mm}
To further analyze the discrepancy between VMC and CCSD(T) for malonaldehyde, we computed the pairwise RMSE between the methods we have tested in this study and plotted these values with a color bar in Fig.~{\ref{fig:pairwise-rmse-heatmap}}. This figure reveals that the consistency between methods differs significantly for ethanol, malonaldehyde, and benzene. For ethanol and benzene, most methods generally show good agreement except for HF and SVWN. However, for malonaldehyde, many method pairs yield RMSE exceeding 10, indicating significant discrepancy among them. Furthermore, the consistency between CCSD and CCSD(T) also varies considerably among ethanol, malonaldehyde, and benzene. While the RMSE between CCSD and CCSD(T) is small for ethanol and benzene, it becomes very large for malonaldehyde. Interestingly, (unbiased and optimized) VMC, PBE0-D3BJ, $\omega$B97X-D3BJ, and $\omega$B97M-D3BJ show good agreement in RMSEs for all three compounds. These results suggest that, for malonaldehyde, VMC does not provide physically incorrect forces; rather, the parenthesis T correction in the CCSD(T) calculation may have a peculiar influence on the atomic forces.

\vspace{2mm}
Although the reason for the significant deviation in malonaldehyde is not yet fully understood, one possibility is that malonaldehyde may have a significant multi-reference character. For CCSD and CCSD(T) calculations, the so-called D1 and T1 diagnostics~{\cite{T1D1diagnostic}} have been proposed as measures to estimate the degree of the multi-reference character. Empirically, it has been advised to pay attention to the reliability of CCSD and CCSD(T) calculations when D1 exceeds 0.05 or T1 exceeds 0.02~{\cite{T1D1diagnostic}}. The D1 and T1 diagnostics taken from the computed 100 structures for ethanol, malonaldehyde, and benzene are D1 values of 0.023 (Max:0.032, Min:0.020), 0.051 (Max:0.063, Min:0.043), 0.030 (Max:0.031, Min:0.029), and T1 values of 0.0097 (Max:0.0115, Min:0.0093), 0.0162 (Max:0.0179, Min:0.0147), 0.0106 (Max:0.0109, Min:0.0103). The D1 value for malonaldehyde shows an average exceeding the threshold (0.05), raising suspicion of a significant multi-reference character. To further investigate this discrepancy, quantum chemistry calculations would require approaches such as CASSCF, while QMC calculations would require VMC or DMC calculations using a multi-determinant ansatz, as reported by Stoolman et al.~{\cite{slootman2024accurate}}. A broader benchmark, e.g., covering more chemically varied systems and including DMC benchmark, is also a promising future work.

\section{Concluding remarks and future perspectives}
\label{sec:concluding-remarks}
In this paper, we introduce an application of the Lagrangian technique in Variational Monte Carlo (VMC) calculations to get atomic forces and pressures consistent with potential energy surfaces (also known as unbiased forces and pressures). The previous study to get unbiased forces and pressures with the Jastrow Slater determinant ansatz requires 6$N$ times additional Density Functional Theory (DFT) calculations ($N$ refers to the number of nuclei in a system)~{\cite{Nakano2024}}, which hinders the applications of unbiased VMC force/pressure calculations in large molecules/crystal or in generations of training data for machine-learning potentials (MLPs). The formulation we report in this paper replaces the additional 6$N$-times DFT calculations with a single coupled-perturbed Hartree-Fock or coupled-perturbed Kohn-Sham calculation. This improves the feasibility of the VMC force calculations with the conventional Jastrow Slater determinant ansatz.
We focus on the Jastrow Slater determinant ansatz with GTO basis sets in this study, as a simple and practical case. However, the Lagrangian technique is versatile for any other functional forms, such as Jastrow multi-determinant ansatz, and any other basis sets, such as plane-waves, with non-variational parameters. The difference from this study is to derive the corresponding Lagrangian depending on the employed ansatz and basis set.
As an application of our developed method, we demonstrate that the unbiased force evaluation improves not only the consistency but also the accuracy of the VMC forces. We selected three molecules from the rMD17 benchmark set~{\cite{Christensen2020_MLST, Christensen2020_rMD17_Figshare}, ethanol, malonaldehyde and benzene, and compared biased VMC forces, unbiased VMC forces, and forces obtained by Coupled-Cluster Singles and Doubles with perturbative Triples [CCSD(T)]~{\cite{Raghavachari1989_CCSDT}} calculations. We found that the biased VMC forces are far from the CCSD(T) ones, while the unbiased ones give values close to those of the CCSD(T) ones.
The unbiased force evaluation described here, for instance, plays a crucial role in generating high-quality training data for MLPs.

\section*{Code and Data availability}
The {\emph{ab initio}} DFT and QMC packages used in this work, CP2K and TurboRVB, are available from their GitHub repositories [\url{https://github.com/cp2k/cp2k}] and [\url{https://github.com/sissaschool/turborvb}], respectively. 
The data to reproduce the rMD17 benchmark test is available from NIMS Materials Data Repository (MDR) [\url{https://doi.org/10.48505/nims.6164}].

\section*{acknowledgments}
K.N. is grateful for computational resources from the Numerical Materials Simulator at National Institute for Materials Science (NIMS), from Research Institute for Information Technology at Kyushu University under the category of General Projects on the supercomputer GENKAI, and from supercomputer Fugaku provided by RIKEN through the HPCI System Research Projects (Project ID: hp250031).
K.N. acknowledges financial support from MEXT Leading Initiative for Excellent Young Researchers (Grant No.~JPMXS0320220025), from Murata Science and Education Foundation (Grant No.~M24AN006), and from Japan Science and Technology Agency (JST), PRESTO (Grant No.~JPMJPR24J9).
This research was also supported by the NCCR MARVEL, a National Center of Competence in Research, funded by the Swiss National Science Foundation (Grant No. 205602).
K.N. acknowledges Michele Casula (CNRS) for fruitful discussion about VMC force calculations.
K.N. acknowledges the valuable comments on the method the author developed before from Dr.~Emmanuel Giner (CNRS) and Dr.~David Ceperley (UIUC) in a conference.
We thank Guglielmo Mazzola (UZH) for organizing a seminar delivered by one of the authors K.N. at UZH. The seminar and subsequent discussions led to this collaboration.
The molecular and crystal structures were depicted using VESTA~{\cite{2011MOM}}.
We acknowledge the use of Grammarly to help us check English grammar.

\appendix
\section{Details of the Variational Monte Carlo calculations}\label{app:vmc}

\subsection{Variational Monte Carlo Methods}
The expectation value of the energy for a given wavefunction, $\Psi$, is evaluated as:
\begin{equation}
\left\langle E \right\rangle  = \frac{{\int {d{\mathbf{x}}{\Psi ^2}\left( {\mathbf{x}} \right) \cdot \hat {\mathcal{H}} \Psi \left( {\mathbf{x}} \right)/\Psi \left( {\mathbf{x}} \right)} }}{{\int {d{\mathbf{x}}{\Psi ^2}\left( {\mathbf{x}} \right)} }} = \int {d{\mathbf{x}}{E_{\rm{L}}}\left( {\mathbf{x}} \right)\pi \left( {\mathbf{x}} \right)},
\end{equation}
where ${\mathbf{x}} = \left( {{{\mathbf{r}}_1}{\sigma _1},{{\mathbf{r}}_2}{\sigma _2}, \ldots {{\mathbf{r}}_N}{\sigma _N}} \right)$ refers to the $N$ electron coordinates and their spins, and 
$$
E_{\rm{L}}\left( {\mathbf{x}} \right) \equiv \frac{\hat {\mathcal{H}}\Psi \left( {\mathbf{x}} \right)}{\Psi \left( {\mathbf{x}} \right)}
\quad  \text{and} \quad 
\pi \left( {\mathbf{x}} \right) \equiv { \frac{{\Psi ^2}\left( {\mathbf{x}} \right)}{\int {d{\mathbf{x'}}{\Psi ^2}\left( {\mathbf{x'}} \right)}} }, 
$$ 
are the local energy and the probability distribution, respectively.
The 3$N$-dimensional integration can be stochastically evaluated by generating a set $\left\{ {{{\mathbf{x}}_i}} \right\}$ according to the probability distribution $\pi \left( {\mathbf{x}} \right)$ generated using the Markov-Chain Monte Carlo (MCMC) technique and by averaging the obtained local energies $E_{\rm{L}}\left( {{{\mathbf{x}}_i}} \right)$:
\begin{equation}
{E_{{\text{MCMC}}}} = {\left\langle {E_{\rm{L}}\left( {\mathbf{x}} \right)} \right\rangle _{\pi \left( {\mathbf{x}} \right)}} \approx \frac{1}{M}\sum\limits_{i = 1}^M {E_{\rm{L}}\left( {{{\mathbf{x}}_i}} \right)}.
\label{eq-eval-observable}
\end{equation}
${E_{{\text{MCMC}}}}$ is always associated with an error since it is evaluated by a stochastic sampling. We note that the error bar is inversely proportional to the square root of the number of samplings.

\vspace{2mm}
Suppose $E_0$ is the ground state energy, the \emph{variational theorem} guarantees that the expectation value of $E$ never becomes smaller than $E_0$. In other words, to approach the ground state, one can optimize a given WF by introducing a set of parameters $\mathbf{p} \equiv \left( \cdots, p_{i}, \cdots \right)$ to the WF ${\Psi \left( {{\mathbf{x}},\mathbf{p} } \right)}$:
\begin{equation}
{E_{{\text{VMC}}}}\left( \mathbf{p}  \right) = \int {d{\mathbf{x}}{E_{\rm{L}}}\left( {{\mathbf{x}}, \mathbf{p}} \right)\pi \left( {{\mathbf{x}}, \mathbf{p}} \right)}  \geqslant E_{0}.
\end{equation}
This framework is called Variational Monte Carlo (VMC).
VMC is conceptually very simple, but the optimization of a many-body WF is a complex numerical task due to several reasons such as the presence of several local minima in the energy landscape and statistical errors associated with MCMC evaluation. Great improvements in this field have been achieved in the QMC community. For instance, in TurboRVB, the automatic differentiation~{\cite{Sorella2010_AAD}} allows the stable and efficient calculation of the energy derivatives with variational parameters and offer an efficient optimization method, the so-called stochastic reconfiguration method (also known as natural gradient optimization)~{\cite{1998SOR, 2007SOR}}.

\subsection{Wavefunction Ansatz}
\label{app:vmc-and-lrdmc}
TurboRVB employs the Jastrow antisymmetrized geminal power (JAGP){~\cite{2003CAS}} ansatz. The ansatz is composed of a Jastrow and an antisymmetric part ($\Psi = J \cdot {\Psi _{{\text{AGP}}}}$). The antisymmetric part reads:
\begin{eqnarray}
{\Psi _{{\text{AGP}}}}\left( {{{\mathbf{r}}_1}, \ldots ,{{\mathbf{r}}_N}} \right) = 
{\det} \left[ {g \left( {{\mathbf{r}}_1^ \uparrow ,{\mathbf{r}}_1^ \downarrow } \right) g \left( {{\mathbf{r}}_2^ \uparrow ,{\mathbf{r}}_2^ \downarrow } \right) \cdots g \left( {{\mathbf{r}}_{N/2}^ \uparrow ,{\mathbf{r}}_{N/2}^ \downarrow } \right)} \right],
\end{eqnarray}
where $g \left( {{\mathbf{r}}_{}^ \uparrow ,{\mathbf{r}}_{}^ \downarrow } \right)$ is called the pairing function. 
Although, for the sake of clarity, an unpolarized system with an even number $N$ of electrons is considered here, it can be generalized to describe a polarized system with unpaired electrons~{\cite{2020NAK2}}.
The pairing function, $g\left( {{{\mathbf{r}}_i^{\uparrow}},{{\mathbf{r}}_j^{\downarrow}}} \right)$, is expanded over Gaussian-type atomic orbitals (GTOs):
\begin{equation}
g\left( {{{\mathbf{r}}_i^{\uparrow}},{{\mathbf{r}}_j^{\downarrow}}} \right) = \sum\limits_{\mu\nu} {P_{\mu \nu}^{\uparrow \downarrow}{\psi_{\mu}}\left( {{{\mathbf{r}}_i^{\uparrow}}} \right){\psi_{\nu}}\left( {{{\mathbf{r}}_j^{\downarrow}}} \right)},
\label{agp_expansion}
\end{equation}
where $\psi(\mathbf{r})$ are primitive or contracted GTOs.
Notice that in Eq.~\ref{agp_expansion}, the expansion coefficients of the pairing functions are labeled with $P_{\mu\nu}^{\sigma\tau}$, to highlight their similarity with the density matrix of Eq.~\ref{eq:density_matrix}.
Depending on the symmetry of ${P_{\mu\nu}^{\uparrow \downarrow}}$, the pairing function is dubbed as AGP with spin-singlet pairs (AGPs), AGP with spin-singlet pairs for triplet correlations (AGPu), and the combination of AGPs and AGPu~{\cite{2020NAK2}}, which correspond to a symmetric, skew-symmetric, and general ${P_{\mu\nu}^{\uparrow\downarrow}}$, respectively~{\footnote{Any square matrix can be uniquely decomposed into a sum of symmetric and skew-symmetric matrices}}.
The basis sets employed in this study are described in the main text. When the JAGP is expanded over {$p$ = $N_{\rm el}/2$} molecular orbitals (i.e., only the occupied molecular orbitals are considered), the JAGP coincides with the Jastrow-Slater determinant (JSD) ansatz~{\cite{2020NAK2}}. In this way, we composed the JSD anstaz, where the SD part was taken from a DFT calculation. Indeed, the coefficients in the pairing functions (i.e., the variational parameters in the SD part) are obtained by CP2K and fixed during the VMC optimization step. Only the coefficients in the Jastrow factor are optimized at the VMC level.

{\vspace{2mm}}
The Jastrow term in this study is composed of inhomogeneous one-body, two-body, and three-body factors ($J = {J_1}{J_2}{J_3}$)~{\cite{2020NAK2}}. We employed a spin-independent Jastrow parametrization in this study. The two-body factor is used to satisfy the electron--electron cusp conditions, and the inhomogeneous one-body and three-/four-body factors are adopted to take the further electron--electron correlation into consideration. The one-body Jastrow factor added to satisfy the electron--nucleus cusp condition is not used in this study because effective core potentials remove the Coulomb singularity.
The inhomogeneous one-body Jastrow part is parameterized as:
\begin{equation}
{J}_1\left( {\mathbf{r}_1}{\sigma _1}, \ldots, {\mathbf{r}_N}{\sigma _N} \right) =  \sum_{i} \sum_{\mu}  {\xi_{\mu} \psi_{\mu}^{\rm{Jas}}\left( {{\mathbf{r}}_i} \right)},
\label{onebody_J_inhom}
\end{equation}
where ${{{\mathbf{r}}_i}}$ are the electron positions, $\xi_{\mu}$ are the variational parameters, and $\mu$ is the GTO index.
The two-body Jastrow factor is defined by a function as:
\begin{equation}
{J_2}\left( {{{\mathbf{r}}_1}, \ldots , {{\mathbf{r}}_N}} \right) = \exp \left( {\sum\limits_{i < j} {v\left( {\left| {{{\mathbf{r}}_i} - {{\mathbf{r}}_j}} \right|} \right)} } \right).
\label{twobody_jastrow}
\end{equation}
$v\left( r \right)$ is a pade-type function:
\begin{equation}
v\left( r \right) = \frac{1}{2}r \cdot {\left( {1 + F \cdot r} \right)^{ - 1}},
\label{twobody_v}
\end{equation}
or the exponential-type function: 
\begin{equation}
v\left( r \right) = \frac{ 1 }{2 F} \left( {1 - {e^{ - F \cdot r}}} \right),
\label{twobody_v_exp}
\end{equation}
where $F$ is a variational parameter.
The three-body Jastrow factors are:
\begin{equation}
{J_{3}}\left( {{{\mathbf{r}}_1}, \ldots , {{\mathbf{r}}_N}} \right) = \exp \left( {\sum\limits_{i < j} {{\Xi _{{\text{Jas}}}}\left( {{{\mathbf{r}}_i},{{\mathbf{r}}_j}} \right)} } \right),
\end{equation}
and
\begin{equation}
{\Xi _{{\text{Jas}}}}\left( {{{\mathbf{r}}_i},{{\mathbf{r}}_j}} \right) = \sum\limits_{\mu, \nu} {\xi_{\mu,\nu}^{}\psi _{\mu}^{{\text{Jas}}}\left( {{{\mathbf{r}}_i}} \right)\psi _{\nu}^{{\text{Jas}}}\left( {{{\mathbf{r}}_j}} \right)},
\label{threebody_jas}
\end{equation}
where $\xi_{\mu, \nu}$ are the variational parameters, and the indices $\mu$ and $\nu$ refer to different orbitals.
The basis sets employed in this study are described in the main text.
The Jastrow factors are optimized using the stochastic reconfiguration method.

\subsection{Computing derivatives}
\label{app:derivatives}
The elements of the right-hand side tensor $\mathbf{B}$ of Eq.~\ref{eq:AZ_eq_B_elements} can be efficiently computed from a VMC sampling:
\begin{eqnarray}\label{eq:fc-vmc-eval}
    B_{\mu i}^\sigma = -2\left\langle{E_{\rm L}(\mathbf{x}) \left[(O_{\mu i}^\sigma(\mathbf{x})-\bar{O}_{\mu i}^\sigma)\right]}\right\rangle \; ,
\end{eqnarray}
where we made apparent the dependence of the local operators on the total electronic coordinate $\mathbf{x}$, sampled by VMC, to distinguish them from constant values. In Eq.~\ref{eq:fc-vmc-eval}, $O_{\mu i}^\sigma(\mathbf{x}) = \partial \ln \Psi(\mathbf{x})/\partial C_{\mu i}^\sigma$, and $\bar{O}_{\mu i}^\sigma \sim \braket{O_{\mu i}^\sigma(\mathbf{x})}$.
We note that ${O}_{\mu i}^\sigma(\mathbf{x})$ can be efficiently computed in a VMC calculation using the adjoint algorithmic differentiation~{\cite{Sorella2010_AAD}}, and the divergences of the generalized forces can be cured by reweighting methods~\cite{Attaccalite2008_AS, Pathak2008_PW_reweight}. It is extremely important that the variational force is evaluated in a covariance form of random variables to reduce its fluctuations~{\cite{2005SOR, 2005UMR}}. 
In practice, TurboRVB always computes the derivatives of $E$ with respect to the AGP matrix elements $P_{\mu\nu}^{\uparrow\downarrow}$ in Eq.~{\ref{agp_expansion}}, $\partial E/\partial P_{\mu\nu}^{\uparrow\downarrow}$. In other words, this is the AO representation of the pairing function.
Thus, the chain rule should be applied to obtain $\partial E/\partial C_{\mu i}^{\sigma}$ from $\partial E/\partial P_{\mu\nu}^{\uparrow\downarrow}$.
Molecular spin orbitals are defined as
\begin{equation}
  \phi_k^{\sigma}(\mathbf r) = \sum_\mu C_{\mu k}^{\sigma}\,\psi_\mu(\mathbf{r}) \; .
  \label{eq:restricted-C}
\end{equation}
Enforcing JSD $\leftrightarrow$ JAGP equivalence (idempotent geminal) via the occupied-orbital expansion,
\begin{equation}
  g(\mathbf r_i^{\uparrow},\mathbf r_j^{\downarrow}) = \sum_{k\in\mathrm{occ}}
  \phi_k^{\uparrow}(\mathbf r_i^\uparrow)\,\phi_k^{\downarrow}(\mathbf r_j^\downarrow) \; ,
  \label{eq:occ-expansion}
\end{equation}
implies the geminal coefficients
\begin{equation}
P_{\mu\nu}^{\uparrow\downarrow} = \sum_{k\in\mathrm{occ}} C_{\mu k}^{\uparrow}\,C_{\nu k}^{\downarrow} \; .
  \label{eq:P-up-dn-from-C}
\end{equation}
and
\begin{equation}
P_{\mu\nu}^{\downarrow\uparrow} = \sum_{k\in\mathrm{occ}} C_{\mu k}^{\downarrow}\,C_{\nu k}^{\uparrow} \; .
  \label{eq:P-dn-up-from-C}
\end{equation}
Define $\tilde B_{\mu\nu}^{\uparrow\downarrow}:=\partial E/\partial P_{\mu\nu}^{\uparrow\downarrow}$ and 
$\tilde B_{\mu\nu}^{\downarrow\uparrow}:=\partial E/\partial P_{\mu\nu}^{\downarrow\uparrow}$.
By the chain rule,
\begin{equation}
  B_{\mu i}^\sigma:=\frac{\partial E}{\partial C_{\mu i}^\sigma}
  = \sum_{\lambda\nu}\frac{\partial E}{\partial P_{\lambda\nu}^{{\uparrow\downarrow}}}
    \frac{\partial P_{\lambda\nu}^{{\uparrow\downarrow}}}{\partial C_{\mu i}^\sigma}
  + \sum_{\lambda\nu}\frac{\partial E}{\partial P_{\lambda\nu}^{{\downarrow\uparrow}}}
    \frac{\partial P_{\lambda\nu}^{{\downarrow\uparrow}}}{\partial C_{\mu i}^\sigma}
  = \begin{cases}
    \sum_{\nu}\tilde B_{\mu\nu}^{\uparrow\downarrow} C_{\nu i}^{\downarrow} + \sum_{\lambda}{\tilde B}_{\lambda\mu}^{\downarrow\uparrow} C_{\lambda i}^{\downarrow} & (\sigma = \uparrow) \\
    \sum_{\lambda}\tilde B_{\lambda\mu}^{\uparrow\downarrow} C_{\lambda i}^{\uparrow} + \sum_{\nu}{\tilde B}_{\mu\nu}^{\downarrow\uparrow} C_{\nu i}^{\uparrow}  & (\sigma = \downarrow).
    \end{cases}
  \label{eq:B-equals-2BtC}
\end{equation}
For spin-restricted cases, $\tilde B_{\mu\nu}^{\uparrow\downarrow}=
\tilde B_{\mu\nu}^{\downarrow\uparrow} :=  \tilde B_{\mu\nu}$, $C_{\nu, i}^{\uparrow} = C_{\nu, i}^{\downarrow} := C_{\nu, i}$, and $\tilde B_{\mu\nu} = \tilde B_{\nu\mu}$ because of the symmetry between spins.
Thus, $B_{\mu i}^\uparrow = B_{\mu i}^\downarrow = 
\sum_{\nu} 2 \tilde{B}_{\mu \nu} C_{\nu i}$.
Notice that we compute the derivatives of the energy with respect to the geminal matrix elements in this study simply because TurboRVB internally holds these values. If one uses a QMC package holding the derivatives of the energy with respect to molecular orbitals, one can just use them for the right-hand side tensor $\mathbf{B}$ of Eq.~\ref{eq:AZ_eq_B_elements}.

\vspace{2mm}
The derivatives of the energy with respect to the atomic positions, i.e., the nuclear forces, with the so-called space warp coordinate transformation (SWCT)~{\cite{Umrigar1989_SWCT}} are computed as~{\cite{Sorella2010_AAD}}:
\begin{eqnarray}
{\bf F}_{\alpha}^{\rm VMC} = - \frac{\text{d}E}{\text{d}{\bf R}_{\alpha}} = &-& \Braket{\frac{\partial E_{\rm L}}{\partial {\bf R}_{\alpha}} + \sum\limits_i^{} {{\omega}\left({\mathbf{R}}_{\alpha},  {{{\mathbf{r}}_i}} \right)\frac{\partial E_{\rm L}}{{\partial {{\mathbf{r}}_i}}}}} \\ &-&2\Braket{(E_{\rm L} - E) \left(\frac{\partial \log \Psi}{\partial {\bf R}_{\alpha}} + \sum\limits_i^{} {\left[ {{\omega}\left({\mathbf{R}}_{\alpha},  {{{\mathbf{r}}_i}} \right)\frac{\partial \log \Psi}{{\partial {{\mathbf{r}}_i}}} + \frac{1}{2}\frac{\partial {\omega} \left( {{{\mathbf{R}}_{\alpha}, {\mathbf{r}}_i}} \right)}{{\partial {{\mathbf{r}}_i}}}} \right]} \right)},
\end{eqnarray}
where
\begin{equation}
{\omega _{\alpha}}\left( {\mathbf{r}} \right) = \frac{{ \kappa \left( {\left| {{\mathbf{r}} - {{\mathbf{R}}_{\alpha}}} \right|} \right)}}{{\sum\nolimits_{{\beta} = 1}^M {\kappa\left( {\left| {{\mathbf{r}} - {{\mathbf{R}}_{\beta}}} \right|} \right)} }},
\end{equation}
with $\kappa\left( r \right) = 1/{r^4}$~\cite{Umrigar1989_SWCT}.
In order to evaluate these differential expressions,  6$N_e$ + 6$N_{\rm a}$ components have to be evaluated, namely, $\left\langle {\cfrac{\partial {E_L}}{{\partial {{\mathbf{r}}_i}}}} \right\rangle$, $\left\langle {\cfrac{\partial \log \Psi}{{\partial {{\mathbf{r}}_i}}} } \right\rangle$, $\left\langle {\cfrac{\partial {E_L}}{{\partial {{\mathbf{R}}_\alpha}}}} \right\rangle $, $\left\langle {\cfrac{\partial \log \Psi }{{\partial {{\mathbf{R}}_\alpha}}}} \right\rangle $~{\cite{Sorella2010_AAD}}. These values are efficiently computed in TurboRVB by using the automatic differentiation technique.

{\vspace{2mm}}
When computing the above derivatives, the so-called reweighting method is essential to mitigate the heavy-tailed distribution problem mentioned in the introduction of the main text. In the MCMC, the probability distribution can also differ from $\pi( {\mathbf{x}} )$. Indeed, one can use an arbitrary probability distribution function $\pi'({\mathbf{x}}) = \Psi _{\text{G}}^2( {\mathbf{x}})/\int {d{\mathbf{x}}\Psi _{\text{G}}^2( {\mathbf{x}} )}$, and estimate a generic local observable $O( {\mathbf{x}} )$ either by using 
$
{{\bar O}_{{\text{MCMC}}}} 
= {\left\langle {O\left( {\mathbf{x}} \right)} \right\rangle _{\pi \left( {\mathbf{x}} \right)}} 
$ 
or:
\begin{equation}
{{\bar O}_{{\text{MCMC}}}} 
= \frac{{{{\left\langle {O\left( {{\mathbf{x'}}} \right)\mathcal{W}\left( {{\mathbf{x'}}} \right)} \right\rangle }_{\pi '\left( {{\mathbf{x'}}} \right)}}}}{{{{\left\langle {\mathcal{W}\left( {{\mathbf{x'}}} \right)} \right\rangle }_{\pi '\left( {\mathbf{x}} \right)}}}} 
\approx \frac{{\sum\nolimits_{i = 1}^{M'} {O\left( {{{{\mathbf{x'}}}_i}} \right)\mathcal{W}\left( {{{{\mathbf{x}}}'_i}} \right)} }}{{\sum\nolimits_{i = 1}^{M'} {\mathcal{W}\left( {{{{\mathbf{x}}}'_i}} \right)} }},
\end{equation}
where $\mathcal{W}\left( {{{{\mathbf{x}}}}} \right) = |{\Psi }\left( {\mathbf{x}} \right)/\Psi _{\text{G}}\left( {\mathbf{x}} \right)|^2$,
and the points $\mathbf{x}'_i$ are distributed according to $\pi'$. This {\it reweighting} scheme is very important when evaluating energy gradients with finite variances.
TurboRVB implements the so-called Attaccalite and Sorella (AS) reweighting scheme~{\cite{Attaccalite2008_AS}} to suppress the divergences of derivatives in the vicinity of the nodal surface. 

{\vspace{2mm}}
\subsection{Error propagation}
\label{app:error-propagation}
Important for error propagation is the fact that the elements $B_{\mu i}^\sigma$ come with an error bar $\delta B_{\mu i}^\sigma$, whereby these elements are also correlated with each other.
In the following, it is most convenient to consider the tensor $\mathbf{B}$ as a vector with a single compound index $B_{\mu i}^\sigma = B_{(\mu i \sigma)} = B_I$, and the elements of $\mathbf{A}$ in Eq.~\ref{eq:AZ_eq_B_elements}, $A_{\mu i,\nu j}^{\sigma\tau} = A_{(\mu i \sigma),(\nu j \tau)} = A_{I,J}$, as forming a matrix of dimension $(N_\text{AO}\times N_\text{MO}\times N_\text{spin})\times (N_\text{AO}\times N_\text{MO}\times N_\text{spin})$.
Let's denote the covariance matrix (tensor) of $X$ as $\operatorname{Cov}(\mathbf{X})$.
With the element-wise notation, Eq.~\ref{eq:AZ_eq_B_elements} is written as:
\begin{align}
  \sum_{\nu j\tau} A_{\mu i, \nu j}^{\sigma\tau} Z_{\nu j}^\tau &= -B_{\mu i}^\sigma \\
  \sum_{\mu i \sigma} (A^{-1})_{\lambda k,\mu i}^{\eta\sigma} \sum_{\nu j\tau} A_{\mu i, \nu j}^{\sigma\tau} Z_{\nu j}^\tau &= -
  \sum_{\mu i \sigma} (A^{-1})_{\lambda k,\mu i}^{\eta\sigma} B_{\mu i}^\sigma \\
  \sum_{\mu i \sigma} \sum_{\nu j\tau} (A^{-1})_{\lambda k,\mu i}^{\eta\sigma}  A_{\mu i, \nu j}^{\sigma\tau} Z_{\nu j}^\tau &= -
  \sum_{\mu i \sigma} (A^{-1})_{\lambda k,\mu i}^{\eta\sigma} B_{\mu i}^\sigma \\
  \sum_{\nu j\tau} \delta_{\nu\lambda} \delta_{jk} \delta_{\eta\tau} Z_{\nu j}^\tau &= -
  \sum_{\mu i \sigma} (A^{-1})_{\lambda k,\mu i}^{\eta\sigma} B_{\mu i}^\sigma \\
  Z_{\lambda k}^\eta &= -\sum_{\mu i \sigma} (A^{-1})_{\lambda k,\mu i}^{\eta\sigma} B_{\mu i}^\sigma \; .
\end{align}
In other words, we have
\begin{align}
  Z_{\mu i}^\sigma &= -\sum_{\nu j \tau} G_{\mu i,\nu j}^{\sigma\tau} B_{\nu j}^\tau
\end{align}
where $G$ is an analogue of the matrix inverse:
\begin{align}
  \sum_{\mu i \sigma} G_{\lambda k,\mu i}^{\eta\sigma}  A_{\mu i, \nu j}^{\sigma\tau} = 
  \delta_{\nu\lambda} \delta_{jk} \delta_{\eta\tau} \; , \quad
  \sum_{\mu i \sigma} A_{\lambda k,\mu i}^{\eta\sigma}  G_{\mu i, \nu j}^{\sigma\tau} =
  \delta_{\nu\lambda} \delta_{jk} \delta_{\eta\tau}
\end{align}
Therefore, $\operatorname{Cov}(\mathbf{Z})$ reads:
\begin{equation}
\mathrm{Cov}(\mathbf{Z})_{\mu  i, \nu j}^{\sigma \tau} =
\sum_{\lambda k \eta,\delta l \zeta}
G_{\mu i, \lambda k}^{\sigma\eta}
\mathrm{Cov}(\mathbf{B})_{\lambda k, \delta l}^{\eta\zeta}
G_{\delta l, \nu j}^{\zeta\tau}.
\end{equation}
These Lagrange multipliers enter the formula for the force only linearly and they get contracted with deterministic quantities only; see Eq.~\ref{eq:lagrangian_derivative}. So, once we obtain $\operatorname{Cov}(\mathbf{Z})$, it is trivial to propagate the error to the forces. Indeed, in Eq.~\ref{eq:lagrangian_derivative}, the force correction term, $F^{\rm{c}}$, can be written as
\begin{equation}
F^{\rm{c}} = const. + \sum_{\mu i \sigma} R_{\mu i}^{\sigma} Z_{\mu i}^\sigma
\; ,
\end{equation}
where $const.$ aggregates all terms that are \emph{not} multiplied by $Z_{\mu i}^\sigma$, and
\begin{equation}
R_{\mu i}^\sigma = \varepsilon_{i \sigma} \sum_{\nu} \frac{\partial S_{\mu \nu}}{\partial {R}} C_{\nu i \sigma} 
+ \sum_{\nu} \frac{\partial F_{\mu \nu \sigma}}{\partial {R}} C_{\nu i \sigma}
- \sum_{jk}\sum_{\lambda\nu\tau} I_{\mu i,jk}^{\sigma\tau} C_{\lambda j}^\tau \frac{\partial S_{\lambda \nu}}{\partial {R}} C_{\nu k}^\tau
\; ,
\end{equation}
with $I_{\mu i,jk}^{\sigma\tau}$ containing the two-electron integrals entering in the definition of $W_{ij}^\sigma$ through the Fock matrix derivative with respect to the MO coefficients, i.e. Eq.~\ref{eq:H_mu_i}.
Thus, the variance of $F^{\rm{c}}$ is
\begin{equation}
\mathrm{Var}(F^{\rm{c}}) = \sum_{\mu i \sigma}\sum_{\nu j \tau}
R_{\mu i}^\sigma \mathrm{Cov}(\mathbf{Z})_{\mu i, \nu j}^{\sigma\tau} R_{\nu j}^\tau \; .
\end{equation}
By plugging $\operatorname{Cov}(\mathbf{B})$ into the above equation, we obtain:
\begin{equation}\label{eq:variance_F_el}
\mathrm{Var}(F^{\rm{c}})
=
\sum_{\mu i \sigma}
\sum_{\lambda k \eta}
\sum_{\delta l \zeta}
\sum_{\nu j \tau}
R_{\mu i}^\sigma
G_{\mu i, \lambda k}^{\sigma\eta}
\mathrm{Cov}(\mathbf{B})_{\lambda k, \delta l}^{\eta\zeta}
G_{\delta l, \nu j}^{\zeta\tau}
R_{\nu j}^\tau \; .
\end{equation}
Dropping the element-wise notation, we get
\begin{equation}\label{eq:variance_F_mat}
\mathrm{Var}(F^{\rm{c}})
=
\vec{R}^\text{T}\mathbf{G}
\mathrm{Cov}(\mathbf{B})
\mathbf{G}^\text{T}
\vec{R} \; .
\end{equation}

\vspace{2mm}
The computation of Eq.~\ref{eq:variance_F_mat} involves the contraction of matrices of dimension
$(N_\text{AO}\times N_\text{MO} \times N_\text{spin}) \times (N_\text{AO}\times N_\text{MO} \times N_\text{spin})$, which results in an overall sixth power with respect to system size. This will limit the maximum system size for which we will be able to compute the error estimation on the forces.
The computation of the $\mathbf{G}$ tensor itself also scales with the sixth power in system size.
A possible way to reduce the computational cost is to perform a low-rank decomposition of the $\mathbf{B}$ tensor:
\begin{equation}
\mathbf{B} = \mathbf{b} \mathbf{b}^\text{T}
\end{equation}
where $\mathbf{b} \in \mathcal{R}^{(N_\text{AO}\times N_\text{MO}\times N_\text{spin})\times N_\text{b}}$, where $N_\text{b}$ is very small. Then, for each of the $N_\text{b}$ vectors, $\mathbf{b}_k$, we have that $\mathbf{b}_k^\text{T} \mathbf{G}$ is the same as solving $\mathbf{A} \mathbf{Z} = \mathbf{B}$, and we can then use a slightly modified version of the force evaluation subroutine to compute $\mathbf{Z}\vec{R}$.

{\vspace{2mm}}
\subsection{Computational details}
The DFT calculations (including the LR calculations) and the VMC calculations in this work were performed on cluster machines operated by the National Institute for Materials Science (NIMS), on the Genkai supercomputer at Kyushu University, and on the Fugaku supercomputer at RIKEN.
For the VMC optimization for Cl$_2$ dimer ,cBN, and AcNH$_2$ test cases reported in Sec.~\ref{sec:validation}, we adjusted the number of independent MCMC chains and the per-chain MCMC sample count so that the error bar of the per-iteration VMC energy estimate was 5.0 mHa. The wavefunction optimization was then run for 300 iterations. As described in the main text, only the Jastrow factor was optimized variationally. Denoting by $\vec{f}$ the vector of the parameter gradients $\equiv \{ \cdots, -\cfrac{\partial E}{\partial p_i} \cdots,\}$, where $p_i$ are the variational parameters, we monitored \verb|devmax| $\equiv \max_i\,\bigl(|f_i|/\lvert\mathrm{std}\,f_i\rvert\bigr)$ and verified for all parameters converged in a statistical sense (\verb|devmax| $\le 4$). The optimization rate and regularization parameters were set to \verb|tpar|=0.35 and \verb|parr|=0.001, respectively. See Ref.~{\onlinecite{2020NAK2}} for the implementation detail in TurboRVB.
After optimization, we performed fresh MCMC simulations to compute the total energy, atomic forces, and the parameter gradients. The energies/forces and parameter gradients were evaluated in separate MCMC runs. For the energy and forces runs, the number of MCMC steps was adjusted so that the error bar of the energy estimate was 0.1 mHa; the resulting error bars on the force components were of comparable magnitude. For the parameter-gradients runs, after equilibration, the total pre-blocking MCMC sample count was set to
$
N_{\mathrm{sample}}
= n_{\mathrm{chain}}\times n_{\mathrm{samples\_per\_chain}}
= 120 \times 30{,}000
= 3{,}600{,}000.
$
For the error bar estimations of the force corrections, we set the number of blocks to 50 per chain.
%
For ethanol, malonaldehyde, and benzene calculations reported in Sec.~\ref{sec:application}, we adjusted the MCMC sample count such that the error bar of the per-iteration VMC energy estimate was 5.0 mHa and performed 100 iterations of VMC optimization. As in Sec.~\ref{sec:validation}, only the Jastrow factor was optimized and \verb|devmax| was confirmed for all parameters. We used \verb|tpar|=0.35 and \verb|parr|=0.001. After optimization, energies, forces, and the parameter gradients were recomputed by MCMC in separate runs. For the energy and force runs, the number of MCMC steps was tuned so that the error bar of the energy estimate was 0.1 mHa, leading to force uncertainties of a similar magnitude. For the parameter-gradient runs, the total number of the pre-blocking MCMC steps was set to
$
N_{\mathrm{sample}}
= n_{\mathrm{chain}}\times n_{\mathrm{samples\_per\_chain}}
= 128 \times 50{,}000
= 6{,}400{,}000.
$
For the calculations in Sec.~\ref{sec:application}, we did not evaluate the error bar of the force-correction term.

{\vspace{2mm}}
\subsection{Parameter study for the rMD17 benchmark test}
\label{app:rmd17-parameter-study}
VMC results are strongly dependent on the chosen ansatz, such as the basis set in the determinant part, the exchange-correlation functional used to prepare the determinant (if it is not optimized at the VMC level), and the functional form of the Jastrow part and its employed basis set. Therefore, we did a benchmark test for many combinations of them. Specifically, we tested cc-pVTZ and cc-pVQZ basis sets accompanied with ccECP for the determinant part, HF calculation and LDA-PZ and B3LYP DFT calculations to prepare the determinant part, the pade and exponential type functional forms for the two-body Jastrow part, and including and not-including the so-called four-body Jastrow parameters (i.e., the pairs of different nuclei in the Jastrow matrix elements) with a  small (H:3s, C:3s1p, O:3s1p), medium (H:3s1p, C:4s2p1d, O:4s2p1d), or large (H:4s2p1d, C:5s3p2d, O:5s3p2d) Jastrow basis set. In this benchmark test set, we used only 20 out of 100 structures.
%
\begin{table}
\centering
\caption{Parameter study for ethanol. For each label, the determinant basis set, ECP, XC, the Jastrow parameters (2-body and 3-/4-body), the Jastrow basis set, and the total number of variational parameters are listed, together with the mean error in energy $\mathrm{ME}_E$ (Ha) and the root-mean-square error in corrected force $\mathrm{RMSE}_F$ (kcal/mol/\AA).}
\label{tab:parameter-study-for-ethanol}
\begin{tabular}{c|cccc|ccc|c|c}
\Hline
Label & Det. Basis & ECP & XC & Jastrow 2b & Jastrow 3b/4b & Jas. Basis & Parameter & ME$_E$ (Ha) & RMSE$_F$ (kcal/mol/\AA) \\
\Hline
1 & cc-pVTZ & ccECP & B3LYP & pade & 3b & small & 148 & +0.4749(19) & 3.451(17) \\
2 & cc-pVTZ & ccECP & B3LYP & pade & 3b/4b & small & 715 & +0.2542(19) & 2.976(17) \\
3 & cc-pVTZ & ccECP & B3LYP & exp & 3b/4b & small & 715 & +0.3480(19) & 3.267(16) \\
4 & cc-pVTZ & ccECP & LDA-PZ & pade & 3b & small & 148 & +0.4443(19) & 3.198(16) \\
5 & cc-pVTZ & ccECP & LDA-PZ & pade & 3b/4b & small & 715 & +0.2296(19) & 2.809(17) \\
6 & cc-pVTZ & ccECP & HF & pade & 3b/4b & small & 715 & +0.4843(18) & 3.711(17) \\
7 & cc-pVTZ & ccECP & B3LYP & pade & 3b & medium & 587 & +0.0832(18) & 2.515(17) \\
8 & cc-pVTZ & ccECP & B3LYP & pade & 3b/4b & medium & 3422 & +0.0387(18) & 2.741(18) \\
9 & cc-pVTZ & ccECP & LDA-PZ & pade & 3b & medium & 587 & +0.0664(19) & 2.344(17) \\
10 & cc-pVTZ & ccECP & LDA-PZ & pade & 3b/4b & medium & 3422 & +0.0(0) & 2.469(19) \\
11 & cc-pVTZ & ccECP & HF & pade & 3b/4b & medium & 3422 & +0.2224(19) & 3.480(20) \\
12 & cc-pVQZ & ccECP & B3LYP & pade & 3b/4b & small & 715 & +0.1811(18) & 3.087(16) \\
13 & cc-pVTZ & ccECP & B3LYP & pade & 3b/4b & large & 13394 & +0.0578(19) & 3.313(27) \\
\Hline
\end{tabular}
\end{table}
%
\begin{table}
\centering
\caption{Parameter study for malonaldehyde. For each label, the determinant basis set, ECP, XC, the Jastrow parameters (2-body and 3-/4-body), the Jastrow basis set, and the total number of variational parameters are listed, together with the mean error in energy $\mathrm{ME}_E$ (Ha) and the root-mean-square error in corrected force $\mathrm{RMSE}_F$ (kcal/mol/\AA).}
\label{tab:parameter-study-for-malonaldehyde}
\begin{tabular}{c|cccc|ccc|c|c}
\Hline
Label & Det. Basis & ECP & XC & Jastrow 2b & Jastrow 3b/4b & Jas. Basis & Parameter & ME$_E$ (Ha) & RMSE$_F$ (kcal/mol/\AA) \\
\Hline
1 & cc-pVTZ & ccECP & B3LYP & pade & 3b & small & 184 & +0.6652(27) & 6.953(24) \\
2 & cc-pVTZ & ccECP & B3LYP & pade & 3b/4b & small & 958 & +0.3877(27) & 7.012(25) \\
3 & cc-pVTZ & ccECP & B3LYP & exp & 3b/4b & small & 958 & +0.5266(28) & 7.525(25) \\
4 & cc-pVTZ & ccECP & LDA-PZ & pade & 3b & small & 184 & +0.6275(28) & 6.953(26) \\
5 & cc-pVTZ & ccECP & LDA-PZ & pade & 3b/4b & small & 958 & +0.3922(27) & 6.974(25) \\
6 & cc-pVTZ & ccECP & HF & pade & 3b/4b & small & 958 & +0.8491(28) & 9.025(28) \\
7 & cc-pVTZ & ccECP & B3LYP & pade & 3b & medium & 803 & +0.0893(27) & 6.886(27) \\
8 & cc-pVTZ & ccECP & B3LYP & pade & 3b/4b & medium & 5069 & +0.0195(28) & 7.213(28) \\
9 & cc-pVTZ & ccECP & LDA-PZ & pade & 3b & medium & 803 & +0.1125(27) & 6.939(27) \\
10 & cc-pVTZ & ccECP & LDA-PZ & pade & 3b/4b & medium & 5069 & +0.0(0) & 6.967(30) \\
11 & cc-pVTZ & ccECP & HF & pade & 3b/4b & medium & 5069 & +0.4373(28) & 8.937(30) \\
12 & cc-pVQZ & ccECP & B3LYP & pade & 3b/4b & small & 958 & +0.2624(29) & 7.946(28) \\
13 & cc-pVTZ & ccECP & B3LYP & pade & 3b/4b & large & 16499 & +0.1427(28) & 7.742(29) \\
\Hline
\end{tabular}
\end{table}

\begin{table}
\centering
\caption{Parameter study for benzene. For each label, the determinant basis set, ECP, XC, the Jastrow parameters (2-body and 3-/4-body), the Jastrow basis set, and the total number of variational parameters are listed, together with the mean error in energy $\mathrm{ME}_E$ (Ha) and the root-mean-square error in corrected force $\mathrm{RMSE}_F$ (kcal/mol/\AA).}
\label{tab:parameter-study-for-benzene}
\begin{tabular}{c|cccc|ccc|c|c}
\Hline
Label & Det. Basis & ECP & XC & Jastrow 2b & Jastrow 3b/4b & Jas. Basis & Parameter & ME$_E$ (Ha) & RMSE$_F$ (kcal/mol/\AA) \\
\Hline
1 & cc-pVTZ & ccECP & B3LYP & pade & 3b & small & 225 & +0.6395(27) & 2.854(18) \\
2 & cc-pVTZ & ccECP & B3LYP & pade & 3b/4b & small & 1548 & +0.4278(27) & 2.494(19) \\
3 & cc-pVTZ & ccECP & B3LYP & exp & 3b/4b & small & 1548 & +0.4589(27) & 2.456(17) \\
4 & cc-pVTZ & ccECP & LDA-PZ & pade & 3b & small & 225 & +0.5619(27) & 2.735(18) \\
5 & cc-pVTZ & ccECP & LDA-PZ & pade & 3b/4b & small & 1548 & +0.3625(27) & 2.518(19) \\
6 & cc-pVTZ & ccECP & HF & pade & 3b/4b & small & 1548 & +0.7913(27) & 2.493(19) \\
7 & cc-pVTZ & ccECP & B3LYP & pade & 3b & medium & 985 & +0.1601(26) & 2.280(18) \\
8 & cc-pVTZ & ccECP & B3LYP & pade & 3b/4b & medium & 8140 & +0.0542(27) & 2.448(19) \\
9 & cc-pVTZ & ccECP & LDA-PZ & pade & 3b & medium & 985 & +0.1288(26) & 2.544(22) \\
10 & cc-pVTZ & ccECP & LDA-PZ & pade & 3b/4b & medium & 8140 & +0.0(0) & 2.091(21) \\
11 & cc-pVTZ & ccECP & HF & pade & 3b/4b & medium & 8140 & +0.3252(26) & 2.480(18) \\
\Hline
\end{tabular}
\end{table}

\vspace{2mm}
As metrics for assessing the quality of the results, we prepared two values. The first one is the VMC energy reported as the value relative to the result for label~10.
For each configuration $i$ in the test set, we define the energy error of the label $x$ relative to the reference (label 10) as
\begin{equation}
\Delta E_i^{({\rm{label}}~x)} = E_i^{({\rm{label}}~x)} - E_i^{({\rm{label}}~10)},
\end{equation}
where $E_i^{({\rm{label}}~x)}$ is the energy computed with the parameter set~(label~$x$) and $E_i^{({\rm{label}}~10)}$
is the reference energy.
The mean error (ME$_E$) is then given by
\begin{equation}
\mathrm{ME}_E^{({\rm{label}}~x)} = \frac{1}{N}\sum_{i=1}^{N}\Delta E_i^{({\rm{label}}~x)},
\end{equation}
where $N$ is the number of configurations ($N = 20$ in this study).
With this notation, $\mathrm{ME}_E > 0$ ($\mathrm{ME}_E < 0$) indicates that label~$x$
gives worse (better) WF than the reference one on average according to the variational principle.
The second one, which is a more direct metric for this study, is the root-mean-square error ($\mathrm{RMSE}$) of all the force components with respect to the reference all-electron CCSD(T) ones computed with the cc-pVQZ basis set.
For each configuration $i$ and force component $\alpha$, we define the force error of a parameter set (label $x$) relative to the reference (all-electron CCSD(T)) as
\begin{equation}
\Delta F_{i,\alpha}^{(\mathrm{label}~x)} =
F_{i,\alpha}^{(\mathrm{label}~x)} - F_{i,\alpha}^{(\mathrm{CCSD(T)})},
\end{equation}
where $F_{i,\alpha}^{(\mathrm{label}~x)}$ is the force component $\alpha$ computed with the parameter set (label~$x$), and $F_{i,\alpha}^{(\mathrm{CCSD(T)})}$ is the reference force.
The RMSE is computed as
\begin{equation}
\mathrm{RMSE}_F^{(\mathrm{label}~x)} =
\sqrt{\frac{1}{N M}\sum_{i=1}^{N}\sum_{\alpha=1}^{M}
\left[\Delta F_{i,\alpha}^{(\mathrm{label}~x)}\right]^2},
\end{equation}
where $N$ is the number of configurations and $M$ is the number of force components ($M = 3$ $\times$ the number of atoms in the system).
The obtained results summarized in Tables~{\ref{tab:parameter-study-for-ethanol}}, {\ref{tab:parameter-study-for-malonaldehyde}}, and {\ref{tab:parameter-study-for-benzene}}. 

\vspace{2mm}
We found, as expected, that the VMC energy decreases as the Jastrow factor becomes more flexible. For example, comparisons between labels~(1 and 2), (4 and 5), (7 and 8), or (9 and 10) indicate that including the four-body term reduces $\mathrm{ME}_E$. However, comparisons such as labels~(2 and 7) or (5 and 9) show that labels~7 and~9 yield smaller $\mathrm{ME}_E$ than labels~2 and~5, respectively, despite having fewer variational parameters. This suggests that using only the three-body term with a larger Jastrow basis is a more efficient way to gain the correlation energy than including a four-body term with a smaller Jastrow basis. For ethanol and malonaldehyde, we also tested an even larger Jastrow basis (label~13). Although some of the 20 configurations exhibited smaller $\mathrm{ME}_E$ than the medium-sized Jastrow basis set (label~8), the average $\mathrm{ME}_E$ became positive. We interpret this as indicating that the optimization becomes less robust for this oversized Jastrow parameterization, probably due to local minima. Therefore, in the main text, we adopt the medium-sized Jastrow basis set.
In contrast to $\mathrm{ME}_E$, the RMSE for force ($\mathrm{RMSE}_F$) does not exhibit a monotonic improvement upon increasing the Jastrow flexibility. For instance, focusing on labels~9 and~10 for the benzene molecule, the inclusion of the 4-body Jastrow improves from 2.544(22)~kcal/mol/\AA~(label~9) to 2.091(21)~kcal/mol/\AA~(label~10), while, for ethanol and malonaldehyde, it does not show a similar improvement. This should reflect the fact that, unlike the total energy, forces are differential quantities (i.e., energy differences with respect to atomic displacements), and improving the wave function does not necessarily improve forces.
On the other hand, there are cases where the choice of the determinant part clearly matters. For example, HF-based calculations (labels~6 and~11) yield substantially worse $\mathrm{RMSE}_F$ than LDA-based calculations (e.g., labels~5 and~10, respectively) for all three molecules. Regarding the basis set for the determinant, we also tested cc-pVQZ for ethanol and malonaldehyde (label~13), but we did not observe improvement in $\mathrm{RMSE}_F$; therefore, we adopt cc-pVTZ in the main text.
We did not find a single parameter set simultaneously yielding the best $\mathrm{ME}_E$ and $\mathrm{RMSE}_F$ for all three compounds. Therefore, we adopt label~10 in the main text because it gives the smallest $\mathrm{ME}_E$ for all three compounds and also provides good $\mathrm{RMSE}_F$ values.

\section{Derivation of Lagrange multipliers equations}\label{app:lagrange}
To arrive at the algebraic equations that determine the Lagrange multipliers, we need to explicitly evaluate the derivatives of the VMC Lagrangian with respect to the MO coefficients and project them onto the occupied and virtual orbital manifolds.
We start by computing the derivatives appearing in Eq. \ref{eq:lagrangian-multiplier}, which we repeat here for convenience:
\begin{eqnarray}\label{eq:app_dL_VMC_dC}
\frac{\partial L_{\rm VMC}}{\partial C_{\mu i}^\sigma} =
\frac{\partial E_{\rm VMC}}{\partial C_{\mu i}^\sigma}
+ \sum_{\lambda k \tau} Z_{\lambda k}^\tau \sum_{\nu} \frac{\partial}{\partial C_{\mu i}^\sigma} 
(F_{\lambda\nu}^\tau C_{\nu k}^\tau - S_{\lambda\nu} C_{\nu k}^\tau \varepsilon_k^\tau)
- \sum_{kl\tau} W_{kl}^\tau \frac{\partial S_{kl}^\tau}{\partial C_{\mu i}^\sigma} \, .
\end{eqnarray}
The first term on the right-hand side is the derivative of the VMC energy with respect to the MO coefficients, which we simply lasbel as $B_{\mu i}^\sigma$ as in the main text.
%
How these derivatives are obtained from the VMC calculation is discussed in Appendix~\ref{app:derivatives}.

{\vspace{2mm}}
The second term in Eq. \ref{eq:app_dL_VMC_dC} corresponds to the derivative of the Roothaan-Hall equations, reading
\begin{align}\label{eq:app_dRH_dC}
\sum_{\lambda k \tau} Z_{\lambda k}^\tau \sum_{\nu} \frac{\partial}{\partial C_{\mu i}^\sigma} 
(F_{\lambda\nu}^\tau C_{\nu k}^\tau - S_{\lambda\nu} C_{\nu k}^\tau \varepsilon_k^\tau) \, .
\end{align}
It is more convenient to work out the action of the derivative operator on the Fock and overlap components separately.
By the chain rule, the first part is given by
\begin{equation}
\begin{aligned}\label{eq:app_dRH_dC_I}
\frac{\partial}{\partial C_{\mu i}^\sigma} F_{\lambda\nu}^\tau C_{\nu k}^\tau &=
\frac{\partial F_{\lambda\nu}^\tau}{\partial C_{\mu i}^\sigma} C_{\nu k}^\tau +
F_{\lambda\nu}^\tau \frac{\partial C_{\nu k}^\tau}{\partial C_{\mu i}^\sigma} \, ,
\end{aligned}
\end{equation}
where the derivative of the Fock matrix with respect to the MO coefficients reads
\begin{equation}
\begin{aligned}\label{eq:app_dF_dC}
\frac{\partial F_{\lambda\nu}^\tau}{\partial C_{\mu i}^\sigma} &=
\frac{\partial}{\partial C_{\mu i}^\sigma} \left(
 h_{\lambda\nu} + \sum_{\gamma\delta\eta} P_{\gamma\delta}^\eta
 [(\lambda\nu|\gamma\delta) - c_x\delta_{\tau\eta}(\lambda\gamma|\nu\delta)]
 + V_{\lambda\nu}^{\text{xc},\tau} \right) \\
&= \sum_{\gamma\delta\eta} \frac{\partial P_{\gamma\delta}^\eta}{\partial C_{\mu i}^\sigma}
[(\lambda\nu|\gamma\delta) - c_x\delta_{\tau\eta}(\lambda\gamma|\nu\delta)]
 + \frac{\partial V_{\lambda\nu}^{\text{xc},\tau}}{\partial C_{\mu i}^\sigma} \\
&= \sum_{\gamma\delta\eta} \left(
\delta_{\sigma\eta} [\delta_{\mu\gamma}C_{\delta i}^\eta + \delta_{\mu\delta}C_{\gamma i}^\eta]
\right) \big[
(\lambda\nu|\gamma\delta) - c_x\delta_{\tau\eta}(\lambda\gamma|\nu\delta)] \big]
 + 2f_{\lambda\nu\tau, \mu i \sigma}^{\rm xc} \\
&= \sum_{\gamma\delta\eta} \delta_{\sigma\eta} \delta_{\mu\gamma} C_{\delta i}^\eta (\lambda\nu|\gamma\delta)
 + \sum_{\gamma\delta\eta} \delta_{\sigma\eta} \delta_{\mu\delta} C_{\gamma i}^\eta (\lambda\nu|\gamma\delta) \\
&- \sum_{\gamma\delta\eta} c_x\delta_{\sigma\eta}\delta_{\tau\eta} \delta_{\mu\gamma} C_{\delta i}^\eta (\lambda\gamma|\nu\delta)
 - \sum_{\gamma\delta\eta} c_x\delta_{\sigma\eta}\delta_{\tau\eta} \delta_{\mu\delta} C_{\gamma i}^\eta (\lambda\gamma|\nu\delta)
 + 2f_{\lambda\nu\tau, \mu i \sigma}^{\rm xc} \\
&= \sum_{\delta} C_{\delta i}^\sigma (\lambda\nu|\mu\delta) + \sum_{\gamma} C_{\gamma i}^\sigma (\lambda\nu|\gamma\mu) \\
&- \sum_{\delta} c_x\delta_{\tau\sigma} C_{\delta i}^\sigma (\lambda\mu|\nu\delta)
 - \sum_{\gamma} c_x\delta_{\tau\sigma} C_{\gamma i}^\sigma (\lambda\gamma|\nu\mu)
 + 2f_{\lambda\nu\tau, \mu i \sigma}^{\rm xc} \\
&= 2(\lambda\nu|\mu i_{\sigma})
 - c_x \delta_{\tau\sigma} [(\lambda\mu|\nu i_{\sigma}) + (\lambda i_{\sigma}|\nu\mu)]
 + 2f_{\lambda\nu\tau, \mu i \sigma}^{\rm xc} \; .
\end{aligned}
\end{equation}
Here, we have introduced the exchange-correlation kernel $f^\text{xc}_{\lambda\nu\tau, \mu i \sigma}$, defined as
\begin{align}\label{eq:app_fxc_kernel}
f_{\lambda\nu\tau, \mu i \sigma}^{\rm xc} = \int \int \frac{\delta^2 E^\text{xc}}
{\delta \rho_\tau (\mathbf{r}) \delta \rho_\sigma (\mathbf{r}')}
\Omega_{\lambda \nu}(\mathbf{r}) \Omega_{\mu i_\sigma}(\mathbf{r}')
\text{d}\mathbf{r}\text{d}\mathbf{r}' \;,
\end{align}
where $\Omega$ denotes an orbital-pair function, $\Omega_{\lambda\nu}(\mathbf r)=\psi_\lambda(\mathbf r)\psi_\nu(\mathbf r)$ (for AOs) and $\Omega_{\mu i_\sigma}(\mathbf r)=\phi_{\mu\sigma}(\mathbf r)\phi_{i\sigma}(\mathbf r)$ (for MOs).
The second term of Eq.~\ref{eq:app_dRH_dC_I} is trivial:
\begin{equation}\label{eq:app_FdC_dC}
F_{\lambda\nu}^\tau\frac{\partial C_{\nu k}^\tau}{\partial C_{\mu i}^\sigma} =
F_{\lambda\nu}^\tau \delta_{\tau \sigma} \delta_{\nu\mu} \delta_{ik} \; .
\end{equation}
Inserting Eqs. \ref{eq:app_dF_dC} and \ref{eq:app_FdC_dC} into Eq. \ref{eq:app_dRH_dC_I} results in
\begin{equation}
\begin{aligned}\label{eq:app_dRH_dC_I_final}
\frac{\partial F_{\lambda\nu}^\tau}{\partial C_{\mu i}^\sigma} C_{\nu k}^\tau +
F_{\lambda\nu}^\tau \frac{\partial C_{\nu k}^\tau}{\partial C_{\mu i}^\sigma} &=
\Bigg[ 2(\lambda\nu|\mu i_{\sigma}) - c_x \delta_{\tau\sigma}
[(\lambda\mu|\nu i_{\sigma}) + (\lambda i_{\sigma}|\nu\mu)]
+ 2 f_{\lambda\nu\tau, \mu i \sigma}^{\rm xc} \Bigg] C_{\nu k}^\tau \\
&+ F_{\lambda\nu}^\tau \delta_{\tau \sigma} \delta_{\nu\mu} \delta_{ik} \; .
\end{aligned}
\end{equation}
The second term of Eq. \ref{eq:app_dRH_dC} is given by:
\begin{equation}\label{eq:app_dRH_dC_II_final}
\begin{aligned}
\frac{\partial}{\partial C_{\mu i}^\sigma} S_{\lambda\nu} C_{\nu k}^\tau \varepsilon_k^\tau
&= 
S_{\lambda\nu} \frac{\partial C_{\nu k}^\tau}{\partial C_{\mu i}^\sigma} \varepsilon_k^\tau
+ S_{\lambda\nu} C_{\nu k}^\tau \frac{\partial \varepsilon_k^\tau}{\partial C_{\mu i}^\sigma} \\
&=
\delta_{\tau \sigma} \delta_{\mu\nu} \delta_{ik} S_{\lambda\nu} \varepsilon_k^\tau
+ S_{\lambda\nu} C_{\nu k}^\tau \frac{\partial \varepsilon_k^\tau}{\partial C_{\mu i}^\sigma}
\; .
\end{aligned}
\end{equation}
Plugging Eqs. \ref{eq:app_dRH_dC_I_final} and \ref{eq:app_dRH_dC_II_final} into Eq. \ref{eq:app_dRH_dC} finally results in:
\begin{equation}\label{eq:app_dRH_dC_final}
\begin{aligned}
\sum_{\lambda k \tau} Z_{\lambda k}^\tau \sum_{\nu}
\frac{\partial(F_{\lambda\nu}^\tau C_{\nu k}^\tau - S_{\lambda\nu} C_{\nu k}^\tau \varepsilon_k^\tau)}
{\partial C_{\mu i}^\sigma}
&= 
\sum_{\lambda k \tau} Z_{\lambda k}^\tau \sum_{\nu} \Big[ C_{\nu k}^\tau \left(
2(\lambda\nu|\mu i_{\sigma}) - c_x \delta_{\tau\sigma} [(\lambda\mu|\nu i_{\sigma}) + (\lambda i_{\sigma}|\nu\mu)]
+ 2f_{\lambda\nu\tau, \mu i \sigma}^{\rm xc} \right) \\
&+
\delta_{\tau\sigma} \delta_{\nu\mu} \delta_{ik} F_{\lambda\nu}^\tau
- \delta_{\tau\sigma} \delta_{\mu\nu} \delta_{ik} S_{\lambda\nu} \varepsilon_k^\tau
- S_{\lambda\nu} C_{\nu k}^\tau \frac{\partial \varepsilon_k^\tau}{\partial C_{\mu i}^\sigma} \Big] \\
&=
\sum_{\lambda \nu k \tau} Z_{\lambda k}^\tau C_{\nu k}^\tau
\left( 2(\lambda \nu | \mu i_{\sigma})
- c_x \delta_{\tau\sigma} [(\lambda\mu|\nu i_{\sigma}) + (\lambda i_{\sigma}|\nu\mu)]
+ 2f_{\lambda\nu\tau, \mu i \sigma}^{\rm xc} \right) \\
&+
\sum_{\lambda} Z_{\lambda i}^\sigma (F_{\lambda\mu}^\sigma - S_{\lambda\mu}\varepsilon_i^\sigma)
- \sum_{k \tau} \underbrace{\sum_{\lambda \nu} Z_{\lambda k}^\tau S_{\lambda\nu} C_{\nu k}^\tau}_{=0}
\frac{\partial \varepsilon_k^\tau}{\partial C_{\mu i}^\sigma} \\
&=
\sum_{\lambda \nu \tau} \tilde{P}_{\lambda\nu}^{\tau} \left( 2(\lambda \nu | \mu i_{\sigma})
- c_x \delta_{\tau\sigma} [(\lambda\mu|\nu i_{\sigma}) + (\lambda i_{\sigma}|\nu\mu)]
+ 2f_{\lambda\nu\tau, \mu i \sigma}^{\rm xc} \right) \\
&+ \sum_{\lambda} Z_{\lambda i}^\sigma (F_{\lambda\mu}^\sigma - S_{\lambda\mu}\varepsilon_i^\sigma) \\
&= H_{\mu i}^\sigma \big[ \tilde{\mathbf{P}} \big] +
\sum_{\lambda} Z_{\lambda i}^\sigma (F_{\lambda\mu}^\sigma - S_{\lambda\mu}\varepsilon_i^\sigma) \; ,
\end{aligned}
\end{equation}
where we have identified the intermediate tensor $\mathbf{H}\big[\tilde{\mathbf{P}}\big]$ as the derivative of the Fock matrix with respect to the MO coefficients (see Eq.~\ref{eq:H_mu_i} in the main text), traced with the asymmetric response spin-density tensor $\tilde{\mathbf{P}}$. Notice that contrary to the main text, here we derive all the equations with the asymmetric response spin-density tensor, though, in practice, we use a symmetrized version as it is numerically more stable.

{\vspace{2mm}}
At last, the third term in Eq.~\ref{eq:app_dL_VMC_dC} is
\begin{equation}
\begin{aligned}
\sum_{kl\tau} W_{kl}^\tau \frac{\partial S_{kl}^\tau}{\partial C_{\mu i}^\sigma} &=
\sum_{kl\tau} W_{kl}^\tau \left( \frac{\partial}{\partial C_{\mu i}^\sigma}
\sum_{\lambda \nu} C_{\lambda k}^\tau S_{\lambda\nu} C_{\nu l}^\tau \right) \\
&= \sum_{kl\tau} W_{kl}^\tau \left(
\sum_{\lambda\nu} \delta_{\lambda\mu} \delta_{ik} \delta_{\tau\sigma} S_{\lambda\nu} C_{\nu l}^\tau +
\sum_{\lambda\nu} \delta_{\mu\nu} \delta_{il} \delta_{\tau\sigma} S_{\lambda\nu} C_{\lambda k}^\tau \right) \\
&= \sum_{kl\tau} \delta_{ik} \delta_{\tau\sigma}  W_{kl}^\tau \sum_{\nu} S_{\mu\nu} C_{\nu l}^\tau +
\sum_{kl\tau} \delta_{il} \delta_{\tau\sigma} W_{kl}^\tau \sum_{\lambda} S_{\lambda\mu} C_{\lambda k}^\tau \\
&= \sum_{l\nu}  W_{il}^\sigma S_{\mu\nu} C_{\nu l}^\sigma +
\sum_{k\lambda} W_{ki}^\sigma S_{\lambda\nu} C_{\lambda k}^\sigma \\
&= 2 \sum_{j\nu} W_{ij}^\sigma S_{\mu\nu} C_{\nu j}^\sigma \; ,
\end{aligned}
\end{equation}
where in the last step we simply relabeled the indices and used the fact that both $\mathbf{W}$ and $\mathbf{S}$ are symmetric.

{\vspace{2mm}}
To determine the Lagrange multipliers, we equate Eq.~\ref{eq:app_dL_VMC_dC} to zero and project it onto the occupied and virtual MO manifolds (see Eqs.~\ref{eq:L_VMC_onto_occ} and \ref{eq:L_VMC_onto_virt} of the main text), solving for $\mathbf{W}$ and $\mathbf{Z}$, respectively. 
Projection onto the virtual manifold results in a linear system of equations, known as the Z-vector equations, while $\mathbf{W}$ drops out. Let us start by explicitly writing out Eq.~\ref{eq:L_VMC_onto_virt}:
\begin{eqnarray}\label{eq:app_dL_VMC_dC_virt}
\sum_\mu \frac{\partial L_{\rm VMC}}{\partial C_{\mu i}^\sigma} Q_{\mu\nu}^\sigma =
\underbrace{\sum_\mu B_{\mu i}^\sigma Q_{\mu\nu}^\sigma}_{\text{term I}} +
\underbrace{\sum_\mu \left[ H_{\mu i}^\sigma[\tilde{\mathbf{P}}] +
\sum_\lambda Z_{\lambda i}^\sigma ( F_{\lambda\mu}^\sigma - S_{\lambda\mu} \varepsilon_i^\sigma ) \right] Q_{\mu\nu}^\sigma}_{\text{term II}} -
\underbrace{\sum_\mu 2 \sum_{j \lambda} W_{ij}^\sigma S_{\mu\lambda} C_{\lambda j}^\sigma Q_{\mu\nu}^\sigma}_{\text{term III}}
= 0 \; ,
\end{eqnarray}
where we recall here the definition of $Q_{\mu\nu}^\sigma$:
\begin{equation}
Q_{\mu\nu}^\sigma = \delta_{\mu\nu} - \sum_{\lambda k} C_{\mu k}^\sigma C_{\lambda k}^\sigma S_{\lambda \nu} \; .
\end{equation}
Term I in Eq.~\ref{eq:app_dL_VMC_dC} remains untouched.
%
Term III vanishes:
\begin{equation}\label{eq:app_term_III_final}
\begin{aligned}
\sum_\mu 2 \sum_{j\lambda} W_{ij}^\sigma S_{\mu\lambda} C_{\lambda j}^\sigma Q_{\mu\nu}^\sigma
&= 2 \sum_{\mu j\lambda} W_{ij}^\sigma S_{\mu\lambda} C_{\lambda j}^\sigma \delta_{\mu\nu}
- 2 \sum_{\mu j\lambda} \sum_{\delta k} W_{ij}^\sigma S_{\mu\lambda} C_{\lambda j}^\sigma
C_{\mu k}^\sigma C_{\delta k}^\sigma S_{\delta\nu} \\
&= 2 \sum_{j\lambda} W_{ij}^\sigma S_{\nu\lambda} C_{\lambda j}^\sigma
- 2 \sum_{j k \delta} W_{ij}^\sigma \underbrace{S_{kj}^\sigma}_{=\delta_{kj}} C_{\delta k}^\sigma S_{\delta\nu} \\
&= 2 \sum_{j\lambda} W_{ij}^\sigma S_{\nu\lambda} C_{\lambda j}^\sigma
- 2 \sum_{j \delta} W_{ij}^\sigma S_{\delta\nu} C_{\delta j}^\sigma \\
&= 0 \; ,
\end{aligned}
\end{equation}
where we have used the fact that the overlap matrix is symmetric.
Finally, we look at term II, which is the most complicated one:
\begin{align}\label{eq:app_term_II}
\sum_{\mu} \left[ H_{\mu i}^\sigma \big[\tilde{\mathbf{P}}\big] 
+ \sum_\lambda Z_{\lambda i}^\sigma (F_{\lambda\mu}^\sigma -S_{\lambda\mu} \varepsilon_i^\sigma)
\right] Q_{\mu\nu}^\sigma &=
\underbrace{\sum_{\mu} H_{\mu i}^\sigma \big[\tilde{\mathbf{P}}\big] Q_{\mu\nu}^\sigma}_{\text{term II.A}}
+ \underbrace{\sum_{\mu\lambda} Z_{\lambda i}^\sigma (F_{\lambda\mu}^\sigma -S_{\lambda\mu}
\varepsilon_i^\sigma) Q_{\mu\nu}^\sigma }_{\text{term II.B}} \; .
\end{align}
Term II.A does not simplify, while 
term II.B simplifies owing to the Roothaan-Hall equation, which vanishes when the Lagrangian is stationary with respect to the MO coefficients:
\begin{equation}\label{eq:app_term_II.B}
\begin{aligned}
\sum_{\mu\lambda} Z_{\lambda i}^\sigma (F_{\lambda\mu}^\sigma -S_{\lambda\mu}
\varepsilon_i^\sigma) Q_{\mu\nu}^\sigma &=
\sum_{\mu\lambda} Z_{\lambda i}^\sigma (F_{\lambda\mu}^\sigma -S_{\lambda\mu}
\varepsilon_i^\sigma) \delta_{\mu\nu} -
\sum_{\mu\lambda} Z_{\lambda i}^\sigma (F_{\lambda\mu}^\sigma -S_{\lambda\mu}
\varepsilon_i^\sigma) \sum_{\delta k} C_{\mu k}^\sigma C_{\delta k}^\sigma S_{\delta \nu} \\
&=
\sum_\lambda Z_{\lambda i}^\sigma (F_{\lambda\nu}^\sigma - S_{\lambda\nu}\varepsilon_i^\sigma)
- \sum_{\lambda\delta k} Z_{\lambda i}^\sigma
\sum_\mu (F_{\lambda\mu}^\sigma C_{\mu k}^\sigma - S_{\lambda\mu}  C_{\mu k}^\sigma  \varepsilon_i^\sigma)  C_{\delta k}^\sigma S_{\delta \nu} \\
&=
\sum_\lambda Z_{\lambda i}^\sigma (F_{\lambda\nu}^\sigma - S_{\lambda\nu}\varepsilon_i^\sigma)
- \sum_{\delta k} \sum_{\lambda\mu} Z_{\lambda i}^\sigma
(S_{\lambda\mu} C_{\mu k}^\sigma \varepsilon_k^\sigma - S_{\lambda\mu} C_{\mu k}^\sigma \varepsilon_i^\sigma) C_{\delta k}^\sigma S_{\delta \nu} \\
&=
\sum_\lambda Z_{\lambda i}^\sigma (F_{\lambda\nu}^\sigma - S_{\lambda\nu}\varepsilon_i^\sigma)
- \sum_{\delta k} \bigg[ \underbrace{\sum_{\lambda\mu} Z_{\lambda i}^\sigma S_{\lambda\mu} C_{\mu k}^\sigma}_{=0} \bigg]
\varepsilon_k^\sigma C_{\delta k}^\sigma S_{\delta \nu} \\
&- \sum_{\delta k} \bigg[ \underbrace{\sum_{\lambda \mu} Z_{\lambda i}^\sigma S_{\lambda\mu} C_{\mu k}^\sigma}_{=0} \bigg]
\varepsilon_i^\sigma C_{\delta k}^\sigma S_{\delta \nu} \\
&=
\sum_\lambda Z_{\lambda i}^\sigma (F_{\lambda\nu}^\sigma - S_{\lambda\nu}\varepsilon_i^\sigma) \; .
\end{aligned}
\end{equation}
Putting together Eq.~\ref{eq:app_term_II} and Eq.~\ref{eq:app_term_II.B}, results in
\begin{align}\label{eq:app_term_II_final}
\sum_{\mu} H_{\mu i}^\sigma \big[\tilde{\mathbf{P}}\big] Q_{\mu\nu}^\sigma
+ \sum_{\mu\lambda} Z_{\lambda i}^\sigma (F_{\lambda\mu}^\sigma -S_{\lambda\mu}
\varepsilon_i^\sigma) Q_{\mu\nu}^\sigma =
\sum_{\mu} H_{\mu i}^\sigma \big[\tilde{\mathbf{P}}\big] Q_{\mu\nu}^\sigma
+ \sum_\lambda Z_{\lambda i}^\sigma (F_{\lambda\nu}^\sigma - S_{\lambda\nu}\varepsilon_i^\sigma) \; .
\end{align}
Finally, inserting Eqs.~\ref{eq:app_term_III_final} and \ref{eq:app_term_II_final} into Eq.~\ref{eq:app_dL_VMC_dC} yields:
\begin{equation}\label{eq:app_dL_VMC_dC_virt_final}
\begin{aligned}
\sum_\mu \frac{\partial L_{\rm VMC}}{\partial C_{\mu i}^\sigma} Q_{\mu\nu}^\sigma
= \sum_{\mu} B_{\mu i}^\sigma Q_{\mu\nu}^\sigma
+ \sum_{\mu} &H_{\mu i}^\sigma \big[\tilde{\mathbf{P}}\big] Q_{\mu\nu}^\sigma
+ \sum_\lambda Z_{\lambda i}^\sigma (F_{\lambda\nu}^\sigma - S_{\lambda\nu}\varepsilon_i^\sigma) = 0 \\
&\Longleftrightarrow \\
\sum_\lambda Z_{\lambda i}^\sigma (F_{\lambda\nu}^\sigma - S_{\lambda\nu}\varepsilon_i^\sigma)
+ \sum_{\mu} &H_{\mu i}^\sigma \big[\tilde{\mathbf{P}}\big] Q_{\mu\nu}^\sigma
= -\sum_{\mu} B_{\mu i}^\sigma Q_{\mu\nu}^\sigma
\; ,
\end{aligned}
\end{equation}
which corresponds to Eq.~\ref{eq:AZ_eq_B_elements} of the main text.

{\vspace{2mm}}
To determine the $\mathbf{W}$ Lagrange multipliers, we project the derivative of the Lagrangian onto the occupied orbitals and equate to zero, cf. Eq.~\ref{eq:L_VMC_onto_occ}. This equation is identical to Eq.~\ref{eq:app_dL_VMC_dC_virt}, the only difference is the projector:
\begin{eqnarray}\label{eq:app_dL_VMC_dC_occ}
\sum_\mu \frac{\partial L_{\rm VMC}}{\partial C_{\mu i}^\sigma} C_{\mu j}^\sigma =
\sum_\mu B_{\mu i}^\sigma C_{\mu j}^\sigma +
\sum_\mu \left[ H_{\mu i}^\sigma[\tilde{\mathbf{P}}] +
\sum_\lambda Z_{\lambda i}^\sigma ( F_{\lambda\mu}^\sigma - S_{\lambda\mu} \varepsilon_i^\sigma ) \right] C_{\mu j}^\sigma -
\sum_\mu 2 \sum_{j \lambda} W_{ij}^\sigma S_{\mu\lambda} C_{\lambda j}^\sigma C_{\mu j}^\sigma = 0 \; .
\end{eqnarray}
No simplifications are possible for the first term, while the third term is simply
\begin{equation}
\sum_\mu 2 \sum_{j \lambda} W_{ij}^\sigma S_{\mu\lambda} C_{\lambda j}^\sigma C_{\mu j}^\sigma
= 2 \sum_{k} W_{ik}^\sigma S_{kj}^\sigma = 2W_{ij}^\sigma \; .
\end{equation}
The middle term simplifies to
\begin{align}
\sum_\mu \left[ H_{\mu i}^\sigma[\tilde{\mathbf{P}}] +
\sum_\lambda Z_{\lambda i}^\sigma ( F_{\lambda\mu}^\sigma - S_{\lambda\mu} \varepsilon_i^\sigma ) \right] C_{\mu j}^\sigma &=
H_{ji}^\sigma[\tilde{\mathbf{P}}] +
\sum_{\mu\lambda} Z_{\lambda i}^\sigma F_{\lambda\mu}^\sigma C_{\mu j}^\sigma
- \underbrace{\sum_{\mu\lambda} Z_{\lambda i}^\sigma S_{\lambda\mu} C_{\mu j}^\sigma}_{=0} \varepsilon_i^\sigma \\
&= H_{ij}^\sigma[\tilde{\mathbf{P}}] +
\underbrace{\sum_{\mu\lambda} Z_{\lambda i}^\sigma S_{\lambda\mu} C_{\mu j}^\sigma}_{=0} \varepsilon_j^\sigma \\
&= H_{ij}^\sigma[\tilde{\mathbf{P}}] \; ,
\end{align}
using the symmetry of $\mathbf{H}$ and Eq.~\ref{eq:Zij_eq_0}.
With these results, we can rearrange Eq.~\ref{eq:app_dL_VMC_dC_occ} to solve for $\mathbf{Z}$, yielding:
\begin{equation}\label{eq:app_Wij_final}
W_{ij}^\sigma = \frac{1}{2}H_{ij}^\sigma[\tilde{\mathbf{P}}] +
\frac{1}{2}\sum_\mu B_{\mu i}^\sigma C_{\mu j}^\sigma \; ,
\end{equation}
that is reported as Eq.~\ref{eq:Wij} in the main text.

\clearpage
\bibliographystyle{apsrev4-2}
\bibliography{./references.bib}

\end{document}